%% file: h2paper2v10mnras.tex
\documentclass[a4paper,usenatbib]{mn2e}
\usepackage{txfonts}
\pdfoutput=1
\usepackage{graphicx,ifthen,url,float,setspace}
\input{xavier_defs.tex}

\def\perd{\;\;\; .}

\newcommand{\nhi}{$N_{\rm HI}$}

\newcommand{\Nth}{2 \sci{20} \cm{-2}}

\newcommand{\ciistr}{C\,II$^*$}
\newcommand{\dla}{DLA}
\newcommand{\dlas}{DLAs}
\newcommand{\htwo}{H$_{\rm 2}$}

\newcommand{\hi}{H\, I}

\newcommand{\expected}{8.56}
\newcommand{\probltone}{0.18}
\newcommand{\expectedtwosigma}{7.60}
\newcommand{\probltonetwosigma}{0.43}

\loadboldmathitalic 
\title [Magellan Uniform Survey of DLAs: Paucity of Strong H$_2$ Absorption]
{The Magellan Uniform Survey of Damped Lyman $\alpha$ Systems  II: Paucity of Strong Molecular Hydrogen Absorption\thanks{This paper includes data gathered with the 6.5 meter Magellan Telescopes located at Las Campanas Observatory, Chile.}}
\author[R.~A. Jorgenson et al.]
{
Regina~A. Jorgenson$^{1,2}$\thanks{NSF Astronomy and Astrophysics Postdoctoral Fellow; raj@ifa.hawaii.edu}, Michael~T. Murphy$^3$, Rodger Thompson$^4$, Robert ~F. Carswell$^2$\\
$^1$ Institute for Astronomy, University of Hawai'i, 2680 Woodlawn Dr, Honolulu, HI 96822, USA \\
$^2$ Institute of Astronomy, University of Cambridge, Madingley Road, Cambridge, CB3 0HA, UK \\ 
$^3$ Centre for Astrophysics and Supercomputing, Swinburne University of Technology, Hawthorn, Melbourne, VIC 3122, Australia \\ 
$^4$ Steward Observatory, University of Arizona, Tucson, AZ 85721, USA \\
}
\date{Submitted \today.}
\begin{document} 
\maketitle 

\begin{abstract}
We present the first large, blind and uniformly selected survey for molecular hydrogen (H$_2$) in damped Lyman $\alpha$ systems (DLAs) with moderate-to-high resolution spectra. 86 DLAs were searched for absorption in the many Lyman and Werner H$_2$ transitions, with $\approx$79\,\% completeness for H$_2$ column densities above $N($H$_2)=10^{17.5}$\,cm$^{-2}$ for an assumed Doppler broadening parameter $b=2$\,km\,s$^{-1}$. Only a single strong H$_2$ absorber was found -- a system detected previously in VLT/UVES spectra. Given our distribution of $N($H$_2)$ upper limits, this $\sim$1\,\% detection rate is smaller than expected from previous surveys at 99.8\,\% confidence. Assuming the $N($H$_2)$ distribution shape from previous surveys, our detection rate implies a covering factor of $\sim$1\,\% for $N($H$_2)\ge10^{17.5}$\,cm$^{-2}$ gas in DLAs ($<6$\,\% at 95\,\% confidence). We obtained new Magellan/MagE spectra for 53 DLAs; 8\,km\,s$^{-1}$-resolution spectra were available for 27 DLAs. MagE's moderate resolution ($\approx$71\,km\,s$^{-1}$) yields weaker $N($H$_2)$ upper limits and makes them dependent on the assumed Doppler parameter. For example, half the (relevant) previous H$_2$ detections have $N($H$_2)\ge10^{18.1}$\,cm$^{-2}$, a factor of just 3 higher than our median upper limit. Nevertheless, several tests suggest our upper limits are accurate, and they would need to be increased by 1.8\,dex to bring our detection rate within 95\,\% confidence of previous surveys.

\end{abstract}

\section{Introduction}\label{introduction}
The damped Lyman $\alpha $ systems (\dlas ), quasar absorption line systems with log \nhi\ $\geq$$\Nth$, currently represent our only probe of normal (i.e. not high mass, highly star forming) galaxies at high redshift.  Also known to dominate the neutral gas content of the Universe from redshift z $\sim$ 5 to today ~\citep{wolfe05}, the \dlas\ likely play a vital role in fueling star formation across cosmic time.   There is much evidence that star formation is occurring in \dlas .  First, \dla\ metallicities tend to be higher than the  Lyman $\alpha$ forest ~\citep{cowie95, wolfe05}, implying either the presence of \textsl{in situ} star formation or gas enriched by previous generations of stars.  Second, the cosmic mean metallicity of \dlas\ appears to evolve with cosmic time ~\citep{pro03met, rafelski12}, evidence for the build-up of metal-enriched gas with cosmic time (though see the caveats due to sample selection bias in ~\cite{rafelski12} and ~\cite{jorgenson13a}).  Third, the indirect measurement of \dla\ star formation rates via the \ciistr\ technique ~\citep{wolfe03a} implies the presence of a modest amount of star formation in at least $\sim$50\,\% of \dlas .  And finally, several \dlas\ have been directly detected in  Lyman $\alpha$, H-$\alpha$ and/or [OIII] emission (i.e. ~\cite{moller04, fynbo10, peroux11, jorgenson14}), direct signatures of the presence of star formation.  

Despite this evidence for star formation in \dlas , actual molecules like molecular hydrogen (\htwo ), a precursor for star formation, are relatively uncommon, and tend to be discovered either serendipitously or by targeting high-metallicity \dlas .  ~\cite{ledoux03} found \htwo\ in 8 out of 33 \dlas , and only a few more by-chance detections have been reported.  Recently,  ~\cite{noter08} compiled observations of 77 primarily archival VLT/UVES \dla\ and sub-\dla\ spectra and found 13 \htwo\ absorbers (12 out of 68 bonafide \dlas ), for a detection rate of $\sim$18\,\%. 

\htwo\ is the primary molecular coolant -- a route through which the energy accumulated by gravitational collapse can be expelled in the form of radiation -- and is thought to be a necessary link in the process of star formation from neutral gas.   The Kennicutt-Schmidt relation ~\citep{kennicutt98} describes the correlation between the gas (molecular plus atomic) mass surface density in a galaxy and its star formation rate per unit area.  Recently, ~\cite{krumholz12} unified observations of star formation rates (SFRs) and their relation to the molecular gas mass density from the small scale molecular clouds of mass 10$^3$$M_{\sun }$ up to the largest submillimeter galaxies with mass $\sim$10$^{11}$$M_{\sun}$, by removing projection effects and found that the star formation rate scales very simply as $\sim$1\,\% of the molecular gas mass density per local free-fall time.  That is, the presence of molecular gas is intimately linked to the process of star formation. The Milky Way and the Magellanic Clouds contain abundant \htwo -- it is detected in most sight-lines through the Milky Way ~\citep{wakker06} and over half of those through the Magellanic Clouds ~\citep{shull00}.  Average \htwo\ fractions, $f \equiv 2 N$(\htwo)$ /[2N$(\htwo)$ + N(\hi )]$, are typically $\sim$10\,\% for the Milky Way and $\sim$1\,\% for the Magellanic Clouds ~\citep{tumlinson02}.  Recently, ~\cite{willingale13} used Gamma Ray Burst X-ray absorption measurements to infer typical Milky Way molecular fractions of $\sim$20\,\%.   
By comparison, the molecular fractions in \dlas\ reported by ~\cite{noter08} are typically much lower, at $f\sim10^{-3}$, but range from $f\sim10^{-1}$ to $10^{-6}$.  

Several authors have attempted to explain this apparent paucity of molecules in the very gas that is purported to turn into stars.  ~\cite{zwaan06} used CO maps of nearby galaxies to determine that 97\,\% of the \htwo\ mass is in systems with N(\htwo) $> 10^{21}$ cm$^{-2}$.  In other words, much of the \htwo\ may simply exist in much higher column density systems than typical \dlas\ which also have small impact parameters, small covering factors and high dust content, making them difficult/rare to observe in quasar surveys.  Other possibilities could include more exotic scenarios such as that found by ~\cite{carswell11} who were motivated by their discovery of a cold, narrow (sub-1 km s$^{-1}$), neutral carbon (C\,I) velocity component to propose the existence of an associated narrow velocity component in \htwo .  Although this narrow component was not apparent in the \htwo\ data because of blending with nearby, broader velocity \htwo\ components, when included in the \htwo\ model fit, the narrow component helped to solve a problem of conflicting radiation fields and \htwo\ level populations in the system (see ~\cite{cui05, carswell11}).  

\begin{figure}
\includegraphics[width=0.99\columnwidth]{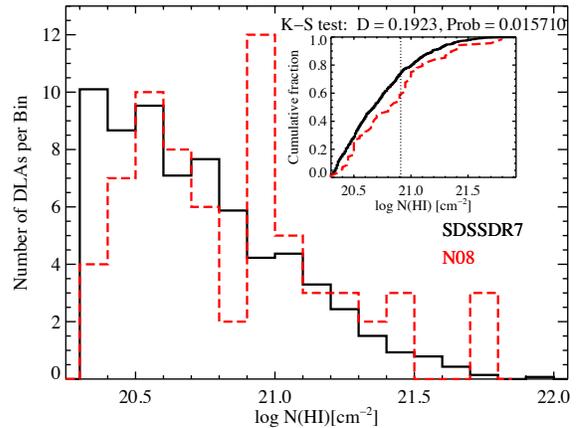}\\ 
\caption{  \nhi\ histogram comparing the Noterdaeme et al. (2008) sample in red (dashed line, labeled `N08') with the scaled SDSS DR7 ~\citep{noter07} distribution in black.  A K-S test indicates the probability they are drawn from the same parent population is P$_{KS}$ $\sim$ 0.016.
}
\label{fig:uveshist}
\end{figure}

Importantly, when discussing the molecular content of \dlas , caution must be employed when interpreting the results of the existing \dla\ surveys for \htwo , i.e. ~\cite{noter08} and ~\cite{ledoux03}, as they consist primarily of a heterogenous mix of archival spectra and \dlas\ that were targeted because of their high-metallicity.  Strong biases can exist in the archival data, i.e. brighter quasars were generally favored for high resolution spectroscopy to obtain a higher signal-to-noise ratio (SNR) with less observing time, and \dlas\ with strong metal lines or high \nhi\ were often targeted as being more interesting.   In Figure~\ref{fig:uveshist} we reproduce the \nhi\ frequency distribution function of the ~\cite{noter08} sample and compare it with that of the SDSS DR7 ~\citep{noter07} distribution.  A K-S test indicates that there is only a $\sim$1.6\,\% chance of these samples being drawn from the same parent population. ~\cite{noter08} attempt to correct for the bias toward large \nhi\ systems in their sample, see their Figure 1, by re-sampling.  However, for the purposes of re-sampling, the sample of 77 \dlas\ and sub-\dlas\ is not large enough and the re-sampling results in large uncertainties.  Recently, ~\cite{balashev14} have searched for \htwo\ in SDSS quasar spectra.  However, given the low resolution (R$\sim$2000) and signal to noise ratio (SNR$\sim$4), they are only able to identify the relatively rare, high column density systems with N(\htwo )$>$10$^{18.5}$cm$^{-2}$.

Until now, no fully blind or untargetted survey for \htwo\ in \dlas\ has been completed and, as a result, the \emph{true}, \emph{unbiased} covering factor and molecular fraction in DLAs remain unknown. A firm understanding of the molecular content in \dlas\ is crucial to understanding the role of these high redshift neutral gas reservoirs in cosmic star formation. 
 In addition, \htwo\ measurements in \dlas\ provide interesting constraints on the physical conditions in the gas, such as temperature, density and ambient radiation fields (e.g. ~\cite{levshakov02, noter07, jorgenson10}), as well as possibly constraining (or detecting) potential variations in the proton-electron mass ratio, $\mu$ (e.g. ~\cite{thompson75, malec10}).  To begin to remedy this situation and to fully understand the connection between \dlas\ and high redshift star formation, we have completed the first large, blind and uniformly selected survey for molecular hydrogen in \dlas\ with moderate-to-high resolution spectra. 

The construction of the survey sample, as well as the \dla\ properties such as metallicity and kinematics are presented in the first paper of this series -- ~\cite{jorgenson13a}.  In this paper, we present the results of the search for \htwo .   
This paper is organized as follows:
In \S~\ref{sec:data} we provide details of the data relevant to the \htwo\ search.  We present our procedure for measuring upper limits on \htwo\ content in each \dla\ in \S~\ref{sec:analysis}.  We then summarize the results of the survey in \S~\ref{sec:results}, comparing them with the results of previous surveys. In \S~\ref{sec:testing} we test the robustness of the routine used to determine upper limits on N(\htwo ) and in \S~\ref{sec:problems} we evaluate the potential impact of systematic errors. Finally, we summarize the results and conclude in \S~\ref{sec:summary}. 

\section{Data and Observations}~\label{sec:data}

We summarize here the details of the sample selection and some issues relevant to the detection of molecular hydrogen.  Details of the observing and data reduction techniques as well as detailed information about each \dla , such as metallicity and low-ion velocity width, are given in Paper 1 ~\citep{jorgenson13a}.  All spectra used in this study are available for download at http://www.dlaabsorbers.info.

\begin{figure*}
\includegraphics[width=1.75\columnwidth]{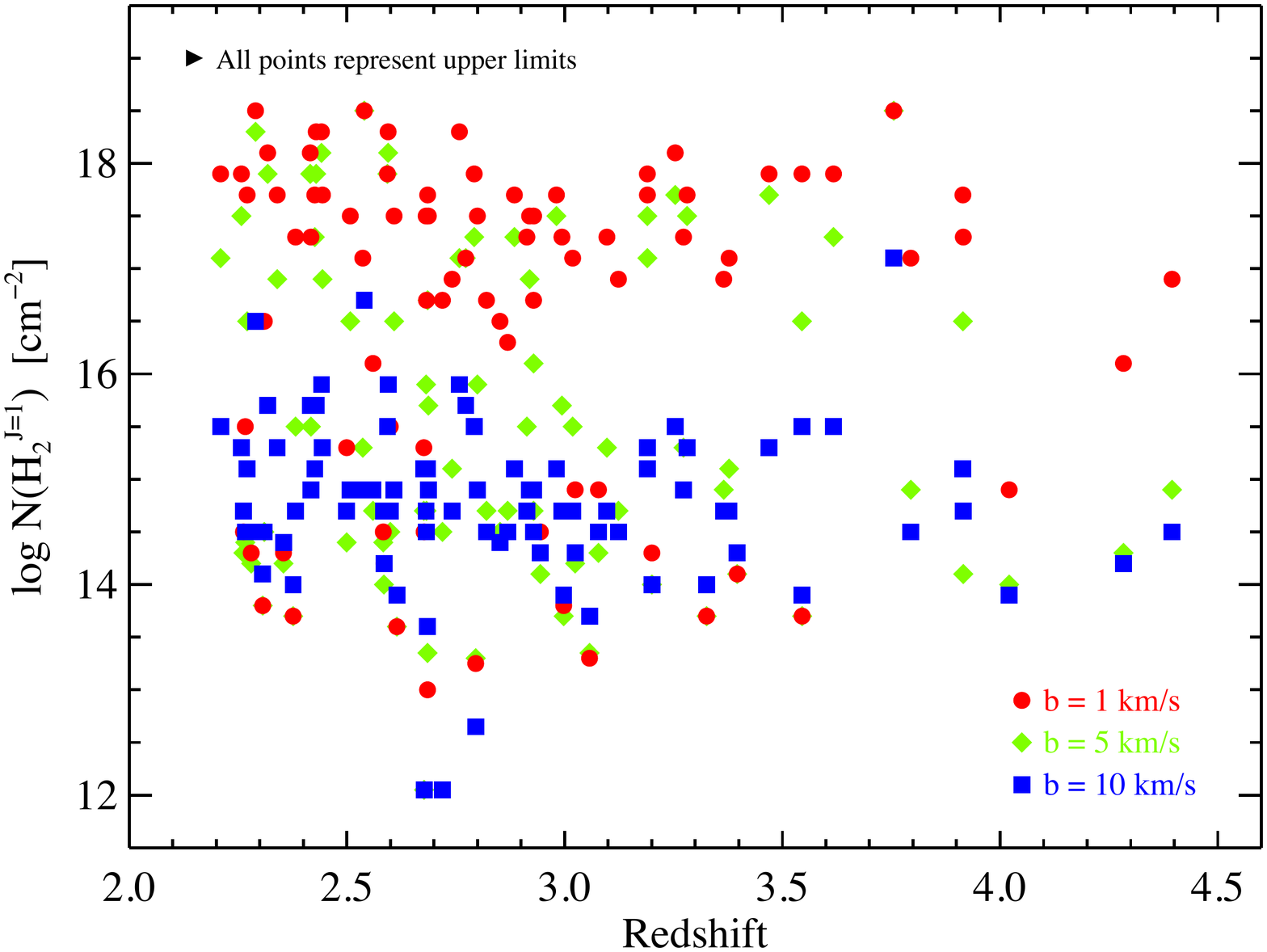}
\caption{Upper limits (1-$\sigma$) on the \htwo\ column density derived from the $J$=1 rotational level versus \dla\ absorption redshift, $z_{abs}$, for 85 \dlas\ for a range of Doppler parameters, $b$ = 1 km s$^{-1}$ (red circles), 5 km s$^{-1}$ (green diamonds), and 10 km s$^{-1}$ (blue squares). This plot gives an idea of the importance of the chosen Doppler parameter in determining the molecular hydrogen column density upper limit.  Upper limits are calculated from the non-smoothed high resolution data when available.   
}
\label{fig:nh2vzallb}
\end{figure*}

\subsection{Sample Selection}~\label{sec:sample}
In constructing the \dla\ sample presented in this paper, our primary goal was to determine the true covering factor and fraction of \htwo\ in \dlas .  To achieve this goal, we created a \dla\ sample drawn with uniform selection criteria from the SDSS DR5 DLA sample of ~\cite{pro05} with the aim of minimizing possible biases. We used just 3 simple selection criteria:  1) the target quasar had to be visible from the Magellan site (dec $\leq\ $ 15$^{\circ}$), 2) the redshift of the \dla\ was required to be $z_{abs}$ $\ge$ 2.2, such that the Lyman and Werner band molecular line region fell at $\lambda^{observed}$ $\ge$ 3200\AA\ and 3) the target quasar had an $i$-band magnitude of $i \leq\ $19, such that we created a reasonably sized sample that could be observed spectroscopically at moderate resolution with non-prohibitive amounts of telescope time.  This selection produced a total of 106 \dlas , towards 97 quasars.  

The resulting \dla\ sample, referred to here as the `Magellan sample,' is unique in the sense that it is the only large, uniformly selected \dla\ survey with medium-resolution (FWHM$\sim$71 \kms\ or higher) spectra allowing for metallicity measurements and detection of strong \htwo\ absorption.  Because our sample was taken directly from the SDSS in an unbiased way, the \hi\ column density distribution, f(\nhi ), is fairly well matched to that of the SDSS DR5 survey (see Figure 1 of ~\cite{jorgenson13a}).  In contrast with other surveys that heavily relied on archival and previously published data to create samples used to measure the \htwo\ fraction ~\citep{noter08, ledoux03} or the cosmic mean metallicity ~\citep{rafelski12, pro03met} of \dlas , the sample presented here was created \emph{a priori} to be an \htwo -blind and independent representation of the \dla\ population without regard to N(HI), metallicity, kinematics, or any other property of the \dla\ system.  

While our sample was designed to be as unbiased as possible, we note that, like any other sample created from the SDSS survey, our sample will contain any of the biases inherent in the SDSS sample. Potentially relevant to the current work is the possibility that the SDSS survey has missed a population of dusty, molecule-rich \dlas\ because excess reddening of the background quasar moved it out of the color selection used for identifying SDSS quasars for spectroscopic follow-up.  While we cannot rule this scenario out completely, we argue that while there is a correlation between \dla\ metallicity and \htwo\ content, there is no trend towards higher \htwo\ content with higher metallicity, as one might expect if there was an even more metal-rich/molecule-rich population. Rather, existing \dla\ samples seem to indicate the existence of a threshold metallicity (of $\approx $ 1/30 solar) above which \htwo\ may form/exist. In addition, it has been shown that dust bias of the magnitude-limited SDSS sample is likely not a major issue (see, for example, ~\cite{ellison01}; ~\cite{murphy04}; ~\cite{jorgenson06}; ~\cite{vladilo08}; ~\cite{frank2010}; ~\cite{khare12}). Finally, we note that our selection criterion of $i \leq\ $19 is not significantly more stringent than that of the SDSS spectroscopic follow-up criterion, which was $i \leq\ $19.1 for $z \lesssim $ 3.0 and $i \leq\ $20.2 for $z \gtrsim $ 3.5 ~\citep{richards02}. 

\input{MNRAS_tab_summ_lls.tex}

\subsection{\htwo\ Sample}
Of the original sample of 106 \dlas\ towards 97 quasars, spectra of 96 \dlas\ towards 88 quasars were obtained either with the Magellan/MagE, or VLT/X-Shooter spectrographs, or from archival VLT/UVES or Keck/HIRES (with some overlap) ~\citep{jorgenson13a}.  Three new \dlas\ were discovered, bringing the total number of \dlas\ to 99. Of these, 13 \dlas\ have been excluded from the sample because of the presence of a higher redshift  Lyman limit system whose Lyman break at 912 \AA\ eliminates the quasar flux in the spectral region containing the associated \htwo\ Lyman and Werner band absorption lines.  Because we are not able to measure nor put an upper limit on the amount of \htwo\ in these systems, they have been taken out of our sample.  Details of these systems are presented in Table ~\ref{tab:lls}. This leaves a remaining 86 \dlas\ towards 77 quasars in our sample to be searched for the presence of \htwo .  

\subsection{Previous \htwo\ detections}
One \dla\ in our sample was already known to contain \htwo , a  $z_{\rm abs}=2.4260$ \dla\ in the line of sight towards SDSS235057.87$-$005209.9 ~\citep{petitjean06}, and hence, we did not re-observe this target with MagE.  ~\cite{noter07} performed a detailed study using a high resolution VLT/UVES spectrum and measured a total N(\htwo ) = 10$^{18.52}$ $\cm{-2}$, and an \htwo\ fraction of $\sim$2\,\%.  While we exclude this \dla\ from the figures and tables consisting entirely of upper limits, we do include this \htwo\ absorber in any statistical results presented. 

\section{Data Analysis}~\label{sec:analysis}

As a `first-pass' search for \htwo , each quasar spectrum was examined by eye to look for signs of obvious, strong molecular absorption lines.  Using the redshift of the absorber determined from an unsaturated, low-ion metal transition (usually the SiII $\lambda$1808 line), we examined the strongest oscillator strength \htwo\ $J$=0 and $J$=1 transitions for consistent absorption signatures.  After this procedure produced no \htwo\ detections we employed a technique for measuring the upper limits on the \htwo\ content in each \dla .

\begin{figure}
\includegraphics[width=0.99\columnwidth]{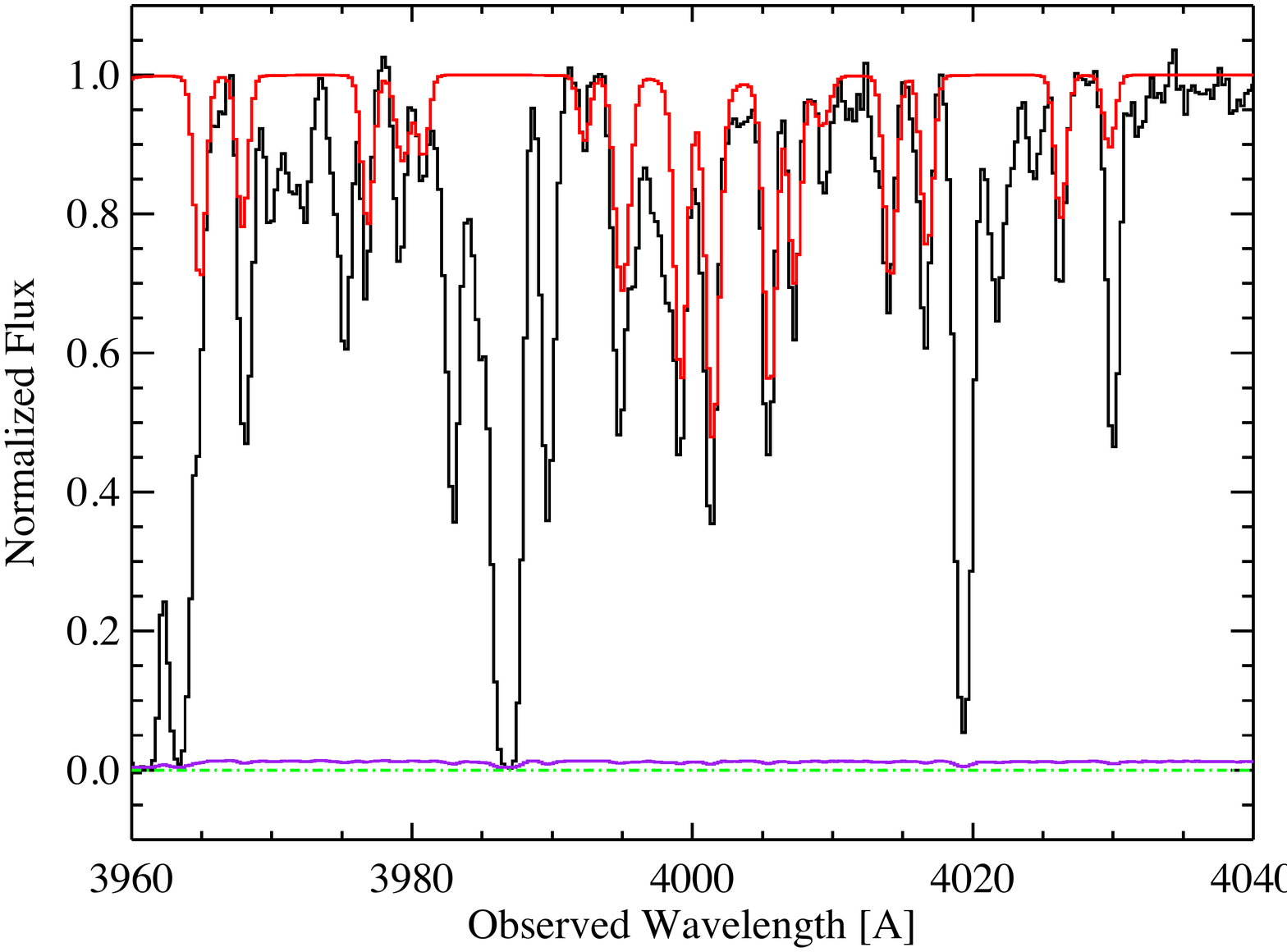}\\
\caption{A section of a Magellan/MagE spectrum of the Lyman $\alpha$ forest region of the quasar PKS 0528$-$250 containing the $z_{abs}$=2.811 \dla .  Overplotted in red is a model fit of the \htwo\ absorption measured in a high resolution VLT/UVES spectrum by Srianand et al. (2005). This spectrum demonstrates that \htwo\ absorption is identifiable in the MagE spectra (with FWHM$\sim$71 km s$^{-1}$).   
}
\label{fig:J0530}
\end{figure}

\input{h2tab_summ_MNRAS1.tex}

\input{h2tab_summ_MNRAS2.tex}

\subsection{Measuring molecular hydrogen upper limits}~\label{sec:upperlims}
Blending with and contamination by the  Lyman $\alpha$ forest is the most significant challenge when measuring the amount of molecular hydrogen contained in a given \dla , and this is a particular problem for lower resolution spectra like those studied here. In order to overcome this problem, we utilize as many of the more than 100 \htwo\ Lyman and Werner band transitions as possible to constrain the measurement of, or put an upper limit on, the total \htwo\ column density, N(\htwo ). 

\begin{figure*}
\includegraphics[width=1.75\columnwidth]{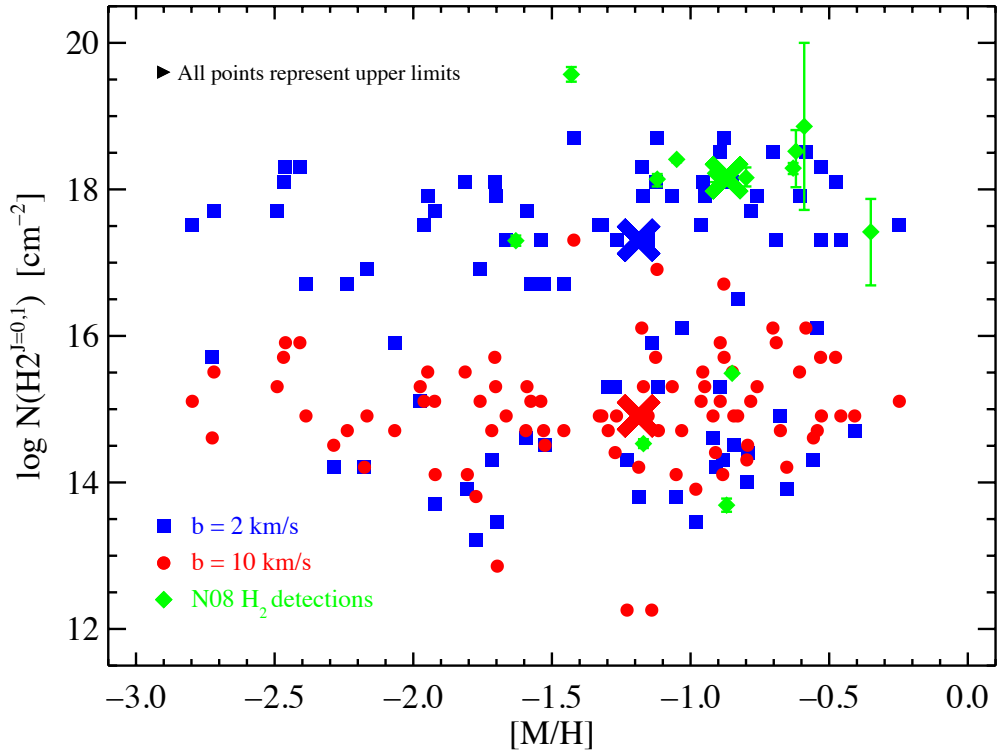}
\caption{Upper limits (1-$\sigma$) on the \htwo\ column density derived from the $J$=0 and 1 rotational levels versus the \dla\ metallicity, [M/H].  Colored points assume various Doppler parameter models as labeled. Green diamonds, labeled `N08,' are the \htwo\ detections 
from the Noterdaeme et al. (2008) sample.  The large `X' represents the median values of the respective samples and is located at the median 
metallicity.
}
\label{fig:nh2vm}
\end{figure*}

We used a routine in the VPFIT \footnote{http://www.ast.cam.ac.uk/$\sim$rfc/vpfit.html} package to determine the upper limits on N(\htwo ).  For a given redshift and Doppler parameter, $b$, a grid of Voigt profiles convolved with the instrument profile is created at every transition of a given \htwo\ rotational $J$ state in the observed spectral range.  The column density of the model fit is then increased until some pixels of the model within $\pm2\sqrt{b_{line}^2 + b_{resolution}^2}$ of the line center (where $b_{resolution} = \frac{\sqrt{2}}{2.3548} \times FWHM_{instrumental} $), have values 1$\sigma$ below the data.  
The $\chi $$^2$ of these pixels is determined and the maximum allowed column density is taken to be that at which the $\chi $$^2$ value over this range had a probability of occurring by chance of less than 0.16, corresponding to a 1$\sigma$, one-sided deviation.  This routine produces a reliable upper limit even in the presence of blends, as it is only those pixels where the model fit is below the data that contribute to the significance level. 

To apply this routine to the sample data, we first trimmed each spectrum to include only sections of spectra over which the median of the signal to noise ratio over a 20 pixel box was greater than or equal to 10\,per 20\,km\,s$^{-1}$ pixel. The SNR degrades uniformly towards the blue and the lowest rest-frame wavelength produced by this SNR-cut represents the wavelength at which the search for \htwo\ in each \dla\ was begun and is given in Table ~\ref{tab:h2sample}. Table~\ref{tab:h2sample} also lists the total wavelength interval, $\Delta$$\lambda$, over which the \htwo\ upper limits were determined.  Given the typical SNR ($\approx$18\,per 20\,km\,s$^{-1}$ pixel) and resolution (FWHM = 71 \kms ) of the spectra, we considered only the strongest \htwo\ transitions, which encompasses all \htwo\ lines of rotational states $J$ = 0, 1, and 2.  We used the wavelengths and oscillator strengths for the \htwo\ transitions provided in ~\cite{malec10}.

Because the VPFIT program treats the column density parameters for each $J$ level independently, the values, or upper limits, for each $J$ state can be physically inconsistent.  Since we are interested in the total \htwo\ we constrained the $J$ = 0, 1 \& 2 transitions by creating a hybrid line list containing all \htwo\ $J$ = 0, 1, 2 transitions, with the oscillator strengths of the $J$ = 0 and $J$ = 2 transitions scaled to the appropriate level assuming a temperature T = 100K.  This approach enabled us to take advantage of additional lines for constraint under the assumption of a reasonable physical model.  

We present the results of our fitting in Table ~\ref{tab:h2sample}, where we summarize the upper limit values for 85 \dlas .  Because the resultant column density strongly depends on the assumed Doppler parameter, we calculated the results for a range of Doppler parameters, $b$ = 1 $-$ 10 km s$^{-1}$.  In Figure~\ref{fig:nh2vzallb} we plot the log N(\htwo ) upper limits of the $J$=1 state (labeled N(\htwo , $J$=1)) for all \dlas\ with sufficient Lyman and Werner band coverage versus the \dla\ absorption redshift of the system for the 85 \dla\ upper limits.  To give an idea of the spread in results and the importance of the assumed Doppler parameter, we plot the upper limits for the range of Doppler parameters, $b$ = 1, 5, and 10 \kms .  The average difference between $b$ = 1 \kms\ and $b$ = 10 \kms\ upper limits is $\approx$1.75 dex.  

Unfortunately, it is generally not possible to accurately predict the \htwo\ Doppler parameter given that of the low-ion metal transitions.  
However, a review of the literature shows \htwo\ $b$-values in the range of 0.5 \kms\ to 19 \kms\ ~\citep{noter08, jorgenson10}, with the majority of \htwo\ $b$-values falling around a few km s$^{-1}$. Averaging over all published \htwo\ Doppler parameters results in a mean $\overline{b}$ = 3.4 \kms\ and median $\langle $ $b$ $\rangle $ = 2.1 km s$^{-1}$.  Therefore, for the remainder of the paper we will present upper limits derived for two cases, $b$ = 2 km s$^{-1}$, as a representative, yet conservative, $b$ -parameter for which meaningful upper limits can be derived and compared to the detections made by N08 and the larger, less conservative Doppler parameter of $b$ = 10 km s$^{-1}$.

\begin{figure}
\includegraphics[width=0.9\columnwidth]{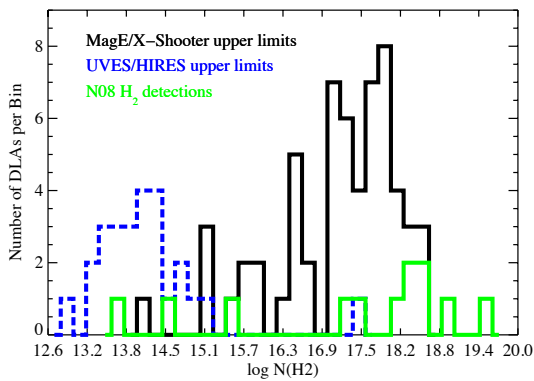}
\caption{Log N(\htwo ) histogram comparing the medium (MagE and X-Shooter, in black) and high (UVES and HIRES, in blue) resolution upper limits of the Magellan sample, assuming $b = 2$ km s$^{-1}$, with the \htwo\ detections of N08 (green).
}
\label{fig:insthist}
\end{figure}

 \begin{figure}
\includegraphics[width=0.99\columnwidth]{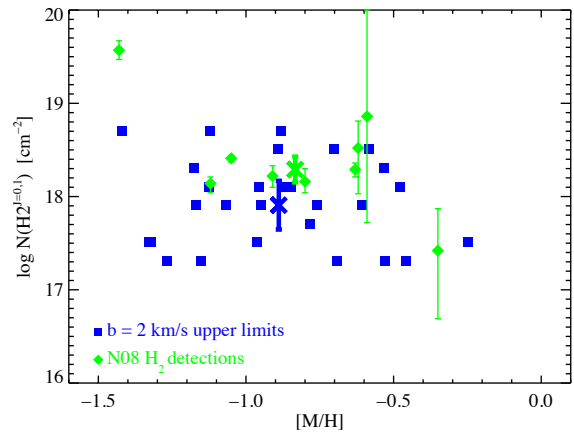}\\
\caption{Zoom in on the previous figure (Fig.~\ref{fig:nh2vm}) plotting 1-$\sigma$ upper limits on the \htwo\ column density derived from the $J$=0 and 1 rotational levels versus the \dla\ metallicity, [M/H].  Green diamonds represent the \htwo\ detections from the Noterdaeme et al. (2008) sample, labeled `N08.'  The large `X' represents the median values of the respective sub-samples, $\langle$log N(\htwo )$^{N08}$$\rangle$ = 18.29 $\pm$ 0.14 (green) and $\langle$log N(\htwo\ )$^{Magellan}$$\rangle$ = 17.91 $\pm$ 0.26 (blue), located at the mean metallicities of each sub-sample.  It is clear that even for the conservative assumption of $b$ = 2 km s$^{-1}$, the \htwo\ upper limits determined from this survey are generally below the majority of the N08 \htwo\ detections.  
}
\label{fig:nh2vm_zoom}
\end{figure}

\subsection{Did we detect known \htwo\ absorption?}~\label{sec:J0530}
As a test of our ability to detect \htwo\ absorption with the medium resolution Magellan/MagE spectrograph (FWHM$\sim$71 km s$^{-1}$), we observed the first \dla\ in which \htwo\ absorption was detected, the \dla\ at $z_{abs}$=2.811 towards PKS 0528$-$250 ~\citep{foltz88}. Note that this \dla\ was not included in our Magellan sample because it was not in the SDSS survey.  In Figure~\ref{fig:J0530}, we plot a section of our Magellan/MagE spectrum of PKS 0528$-$250, in black.  Overplotted in red is a model \htwo\ fit based upon the \htwo\ measurements reported by ~\cite{srianand05} from a high resolution VLT/UVES spectrum.  It is clear that the \htwo\ absorption is easily distinguishable, even at the MagE resolution.

Using VPFIT to calculate the upper limit on the allowed \htwo\ column density following the same procedure used for the MagE spectra of this sample, we find N(\htwo )$^{total}$ $\leq$ 10$^{18.11}$ cm$^{-2}$ at the best-fit redshift, $z_{abs}$= 2.8111 and assuming $b$ = 2 km s$^{-1}$ and temperature T = 100 K for populating the $J$ = 0 and 2 states.  This can be compared with the measured N(\htwo ) from the UVES spectrum, N(\htwo )$^{total}$ = 10$^{18.23}$ cm$^{-2}$.  Given the uncertainties in continuum placement, our assumption of T = 100 K for populating the rotational $J$ levels and the differences in spectral resolution (FWHM$\sim$8 \kms versus 70 km s$^{-1}$), this agreement is quite good. 

Repeating this exercise but now using the detailed velocity structure determined from the high resolution UVES spectrum, we can compare the MagE-determined upper limits with the amount of \htwo\ measured in the UVES spectrum.  Measuring upper limits in the MagE spectrum at both UVES-determined \htwo\ velocity components, $z_{abs}$= 2.81100 and $z_{abs}$= 2.81112  ~\citep{srianand05}, we find N(\htwo )$^{total}$ $\leq$ 10$^{18.26}$ cm$^{-2}$, consistent with the UVES measurement.

\section{Results and Analysis}~\label{sec:results}

\begin{figure*}
\includegraphics[width=1.75\columnwidth]{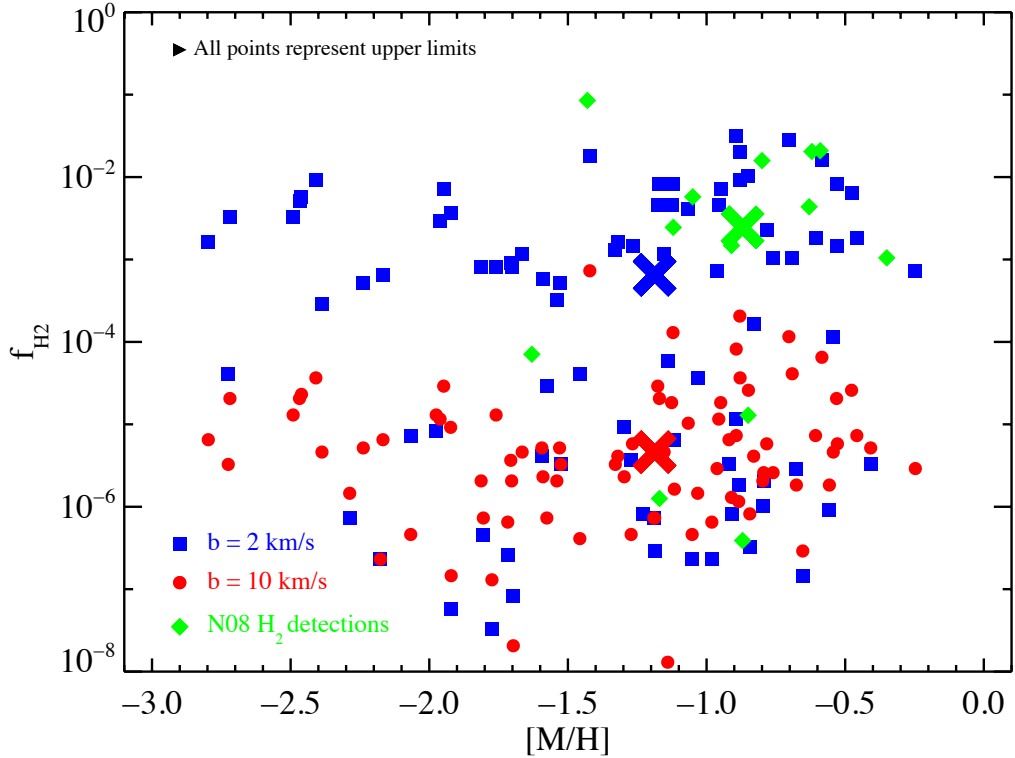}
\caption{Upper limits (1-$\sigma$) on the molecular hydrogen fraction, $f$, versus the \dla\ metallicity, [M/H], for the $b$ = 2 \kms\ (blue squares) and $b$ = 10 \kms\ (red circles) assumed Doppler parameters. Green diamonds represent the \htwo\ detections from the sample of Noterdaeme et al. (2008), labeled `N08.'  The large 'X' represents the median value of the respective sample and is located at the median [M/H]. 
}
\label{fig:fracvm}
\end{figure*}

\begin{figure*}
\includegraphics[width=1.75\columnwidth]{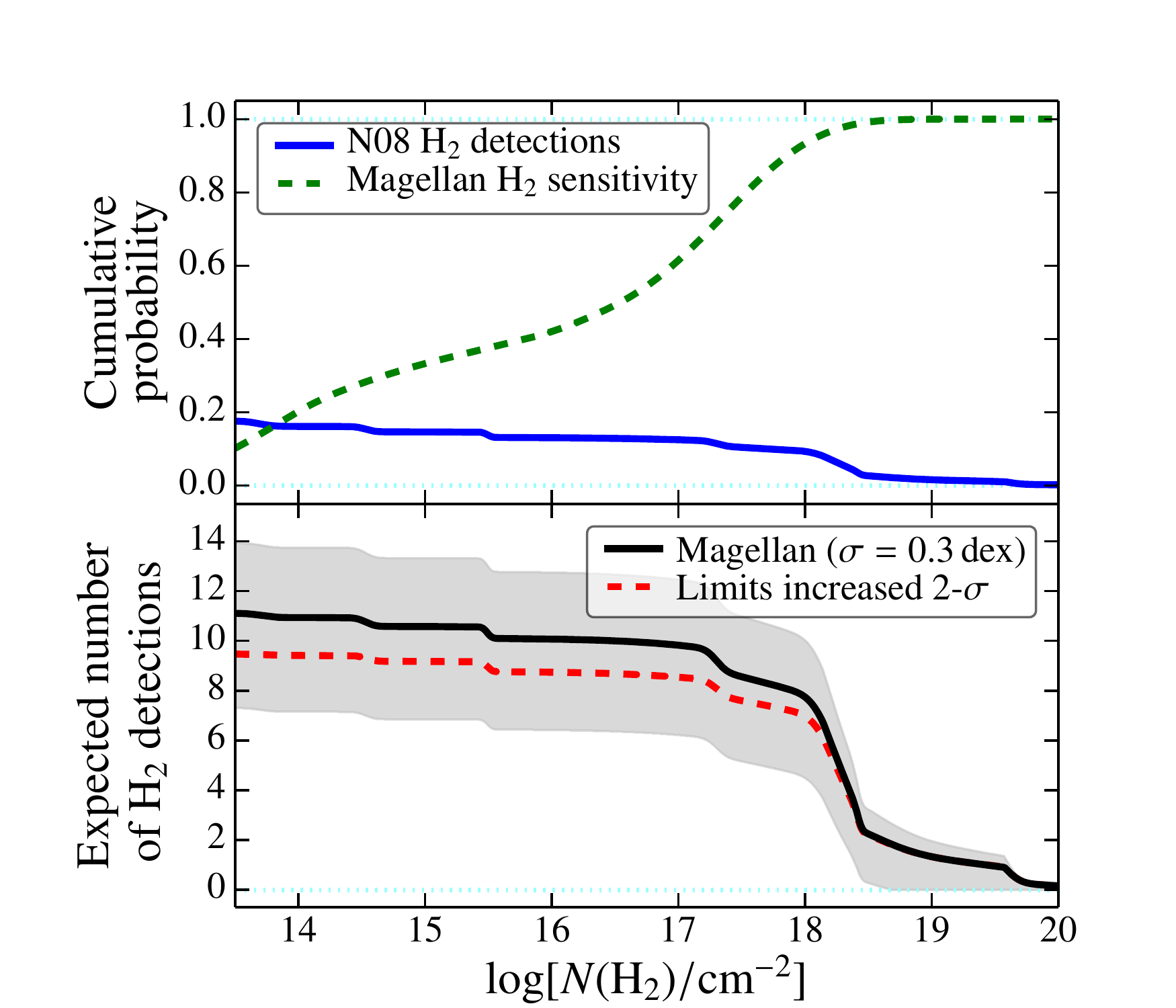}
\caption{{\bf Top Panel:} The cumulative probability distribution of the N08 \htwo\ detections (blue solid line) and the cumulative distribution of the Magellan sample \htwo\ upper limits (green dashed). {\bf Bottom Panel:} The number of \htwo\ detections expected in the Magellan sample as a function of threshold \htwo\ column density (black line) assuming the true \htwo\ distribution is given by the N08 sample.  The grey shaded region defines the 1$\sigma$ confidence region on this expectation assuming Poisson statistics.  The red dashed line indicates the expected distribution assuming the upper limits of the Magellan sample have been underestimated by 2$\sigma$ (i.e. if the upper limits are increased by 0.6 dex).     
}
\label{fig:detprob}
\end{figure*}

The major result of this survey is that we did not unambiguously detect any new, strong \htwo\ absorbers in our sample of 86 \dlas . We show below that this result leads to the conclusion, assuming that the \htwo\ upper limits are non-detections, that the covering factor of \htwo\ for N(\htwo ) $\geq\ 10^{17.5}$ cm$^{-2}$ in \dlas\ is $\approx$ 1 \,\%, significantly lower than the estimates of previous surveys ~\citep{noter08, ledoux03}.

To place the results of this survey for molecular hydrogen in context, we will compare our results with those of the survey for \htwo\ in \dlas\ by ~\cite{noter08}, which includes as a subset the previous survey by ~\cite{ledoux03}.  The ~\cite{noter08} \htwo\ survey, hereafter also referred to as the `N08 sample,' consists of 77 absorbers with \nhi\ $\geq$ 10$^{20.0}$ $\cm{-2}$ (of these 68 are \dlas\  with \nhi\ $\geq$ 10$^{20.3}$ $\cm{-2}$ ) in primarily archival VLT/UVES data.  They detect \htwo\ in 13/77, or $\approx$17\,\% of their sample (12/68 or $\approx$18\,\% if only considering bonafide \dlas ) with a wide range of H$_2$ column densities [N(H$_2$)$ = 10^{13.69} - 10^{19.57}$ cm$^{-2}$] and molecular fractions, $f$ $\equiv$ 2 N(\htwo )/(2 N(\htwo ) $+$ N(\hi )), from $f \simeq$ 5 $\times$ 10$^{-7}$ to $f \simeq$ 0.1 ~\citep{noter08}. 

In order to compare with the results reported in ~\cite{noter08}, we plot in Figure~\ref{fig:nh2vm} the upper limits on the total N(\htwo ) contained in the $J$=0 and $J$=1 rotational states, denoted log N(\htwo $^{J=0,1}$), for 85 of our sample \dlas , versus the \dla\ metallicity as derived in ~\cite{jorgenson13a}.  For clarity, we plot only the upper limits obtained assuming the Doppler parameters $b$ = 2 and 10 km s$^{-1}$.  The results from the N08 sample, considering only the detections for clarity, are overplotted as green diamonds for comparison.  We do not present the upper limits from the N08 sample because it was not immediately clear whether they were determined with a method comparable to our method using VPFIT.  The median values of each sample are denoted by the large `X' and are N(\htwo $^{J=0,1}$) = 10$^{18.16}$ $\cm{-2}$, 10$^{17.31}$ $\cm{-2}$ and 10$^{14.91}$$\cm{-2}$ for the N08, $b$ = 2 km s$^{-1}$ and $b$ = 10 km s$^{-1}$ samples, respectively.  It is clear that even under the conservative assumption of $b$ = 2 km s$^{-1}$, the \htwo\ upper limits of our blind survey for \htwo\ in our uniformly selected quasar observations are generally low enough to have detected the known \htwo\ absorbers reported in the N08 sample. 

In comparing the Magellan sample with that of N08 we note that the Magellan sample contains spectra from a variety of instruments. While the bulk of our survey consists of Magellan/MagE spectra with FWHM$\sim$71 km s$^{-1}$ and a median SNR$\sim$18 pixel$^{-1}$, higher resolution spectra from VLT/UVES and Keck/HIRES with FWHM$\sim$8 km s$^{-1}$ were available for 27 of the 86 sample \dlas , and 6 from VLT/X-Shooter (FWHM$\sim$59 km $^{-1}$).  We plot the distribution of upper limits, assuming $b = 2$ km s$^{-1}$, in Figure~\ref{fig:insthist} and compare with the N08 N(\htwo ) detections.  It is clear that the upper limits derived from the medium resolution MagE and X-Shooter spectra are higher, as expected, than those of the UVES and HIRES spectra.  While several of the medium resolution upper limits are near the bulk of the N08 detections, we discuss in Section~\ref{sec:testing} how many of these upper limits are relatively high because of limited wavelength coverage and/or low SNR, rather than being due purely to lower resolution.  We note here that if we restrict the sample to only MagE spectra, the median upper limit is N(\htwo ) = 10$^{17.7}$ cm$^{-2}$ (assuming $b =$ 2 km s$^{-1}$), still smaller than that of N08, N(\htwo ) = 10$^{18.16}$ cm$^{-2}$.

We next construct subsets of the samples consisting of only the high metallicity, high N(\htwo) detections (or limits).  This region, shown in  Figure~\ref{fig:nh2vm_zoom} and taken to be N(\htwo ) $\ge$ 10$^{17}$ $\cm{-2}$ and [M/H]$\geq -$1.5, is the region in which the Magellan survey $\emph{should}$ have detected \htwo \ if it were present.  It can be seen by eye that even in this region the upper limits of the Magellan survey are generally less than the detections of the N08 sample.  A K-S test gives the probability that these two sub-samples are drawn from the same parent population, P$_{KS}$ = 1.1 $\times$ 10$^{-5}$.  The median values of the two samples are denoted by the large blue and green `X' for the Magellan sample upper limits ($\langle$ N(\htwo\ )$^{Magellan}$$\rangle$ = 10$^{17.91\pm 0.26}$  $\cm{-2}$) and the N08 detections ($\langle$ N(\htwo\ )$^{N08}$$\rangle$ = 10$^{18.29 \pm 0.14}$ $\cm{-2}$). We note that the upper limits of the Magellan sample are determined using the conservative Doppler parameter $b$ = 2 \kms\ and, if the true Doppler parameters are larger, it would imply a decrease in the \htwo\ column density upper limits and increase the expectation of detecting new \htwo\ absorbers in the Magellan sample.

  We find similar results if we instead examine the molecular hydrogen fraction, $f \equiv 2 N$(\htwo)$ /[2N$(\htwo)$ + N(\hi )]$.
 The associated molecular fraction upper limits are shown in Figure~\ref{fig:fracvm}, where the median values, denoted by `X', are $f$ = 2.45 $\times$10$^{-3}$, 6.51 $\times$10$^{-4}$ and 4.61 $\times$10$^{-6}$ for the N08, $b$ = 2 km s$^{-1}$, and $b$ = 10 \kms\ samples, respectively.

\subsection{Expected \htwo\ content}~\label{sec:expectedh2}

Assuming that the N08 sample represents the true molecular content in \dlas\  (ignoring the previously mentioned issues with sample biases), what is the likelihood that the Magellan sample would recover a detection rate of $\sim$1\% by chance?  To answer this question and to present a fair comparison between the samples, we must compare the \htwo\ detection rates within the regime where both surveys are sensitive.  We note that at the minimum SNR used to search for \htwo\ (SNR = 10, with 2 exceptions), the Magellan survey is generally sensitive to detecting \htwo\ at the level of N(\htwo ) $\gtrsim$ 10$^{18}$ cm$^{-2}$, assuming $b = $ 2 km s$^{-1}$  (see Section~\ref{sec:sens}).  Therefore, we will compare the Magellan sample with the N08 sample detection rate of `strong' molecular hydrogen systems, defined here as those systems with N(\htwo ) $\geq$ 10$^{18}$cm$^{-2}$.  From N08 we find 7 out of 68 \dlas\ with N(\htwo ) $\geq$ 10$^{18}$cm$^{-2}$, for a detection rate of $10.3^{+3.1}_{-4.6}$\,\%, using Poisson errors. Given this probability distribution, there is a 0.39\% chance that the Magellan sample, considering only the 75 upper limits with N(\htwo ) $\leq 10^{18}$ cm$^{-2}$,  would recover a detection rate of 1\% or less.  In other words, with this crude approach, the detection rate of strong \htwo\ systems in the Magellan sample is smaller than that of the N08 sample at the 99.61\% confidence level.  

A more detailed treatment, the results of which are summarized in Figure~\ref{fig:detprob}, is as follows. We have calculated the number of \htwo\ detections expected in the Magellan sample at a given \htwo\ column density threshold by assuming that our 1-$\sigma$ upper limits will, 68\,\% of the time, sit 0.3\,dex above the most likely column densities; that is, we assume $\sigma=0.3$\,dex.  This is a reasonable and conservative estimate given the tests discussed in Section~\ref{sec:testing}.   We then used the error function to construct the probability of detecting a given \htwo\ column density or higher, the \htwo\ detection function, $D_i(x)$,  for DLA $i$ at a logarithmic column density $x=\log[N$(H$_2$)$]$. If we assume that the probability density function of of \htwo\ column densities, $k_{\rm N08}$, is given by the N08 sample, we can then calculate the probability of a given \htwo\ column density existing and it being detected in the Magellan sample for a given \dla\ as,

\begin{equation}
P^{det}_{i}(x) = \int _x^\infty k_{N08}  D_i(x) \ dx \perd
\end{equation}

\noindent We then find the number of \htwo\ detections that we expect at a given $x=\log[N$(H$_2$)$]$ or higher to be

\begin{equation}
N^{det}(x) = \sum _i^{N_{DLA}} P^{det}_{i}(x)\perd
\end{equation}

\noindent We plot $N^{det}$ as a function of log[N(\htwo )] in the bottom panel of Figure~\ref{fig:detprob}.  It is seen that at an \htwo\ column density of log[N(\htwo )/cm$^{-2}$] = 17.5, where the Magellan sample should have 79\,\% overall sensitivity to detecting \htwo , the expected number of \htwo\ systems is \expected .  The probability of detecting $\le$1 \htwo\ system in the Magellan sample is \probltone \% (assuming Poisson statistics) at this column density threshold of log[N(\htwo )/cm$^{-2}$] $\geq 17.5$.  We conclude from this exercise that we detect too few \htwo\ absorbers with high column densities, N(\htwo ) $\ge$ 10$^{17.5}$ cm$^{-2}$, if we assume that the N08 survey results define the logN(\htwo) distribution in an unbiased way.  

Using this same formalism we can estimate the expected covering factor of \htwo\ for a given \htwo\ column density threshold or higher by changing the normalization of the $k_{N08}$ function until it is consistent with the results of the Magellan survey.  That is, we assume the same shape of $k_{\rm N08}$ but change the relative normalization above the chosen column density threshold of log[N(\htwo )/cm$^{-2}$] $\geq 17.5$.  In the first case we suppress the normalization of $k_{N08}$ by a factor of 1.75 above the threshold such that the detection of a single absorber, as in the Magellan sample, is consistent within 2$\sigma$ of the expected number of detections at N(\htwo ) $\ge$ 10$^{17.5}$ cm$^{-2}$.  This leaves 6\,\% of the total \htwo\ column density distribution above the threshold.  That is, the 2$\sigma$ upper limit on the covering factor of N(\htwo ) $\ge$ 10$^{17.5}$ cm$^{-2}$ gas from the Magellan survey is 6\, \%.  Similarly, if instead we demand consistency with the detection of 1 \htwo\ system, as found in the Magellan survey, we find that we must suppress the normalization of $k_{N08}$ by a factor of 8.5 such that the covering factor of \htwo\ is 1.2\%.  Therefore, the most likely covering factor implied by the Magellan sample, given the above assumptions, for N(\htwo ) $\ge$ 10$^{17.5}$ cm$^{-2}$ gas in \dlas\ is 1.2\% with a 95\% confidence limit of 6\%.   

 If for some reason we have systematically underestimated the upper limits of the Magellan sample, the above formalism allows a simple test of the effect on our conclusions.  For example, if we increase the Magellan upper limits by 2$\sigma$, i.e. by 0.6 \,dex, we find that the number of expected detections with N(\htwo) $\ge$ 10$^{17.5}$ cm$^{-2}$ is \expectedtwosigma\ (with Posisson probability of $\le$1 of \probltonetwosigma \%), seen from the red dashed line in Figure~\ref{fig:detprob}.  If we instead ask by how much we must have underestimated our upper limits for our data to be consistent with N08, we find that we would need to increase the Magellan upper limits by 6$\sigma$, or 1.8 dex, in order to bring our detection of a single N(\htwo) $\ge$ 10$^{17.5}$ cm$^{-2}$ system into 95\% confidence agreement with N08.  There is no evidence to support such a systematically large underestimate in the upper limits derived for the Magellan sample and thus, this seems an unlikely explanation for the different detection rates and \htwo\ covering factors found between the two surveys.

\section{Analysis testing: The Upper limits routine}~\label{sec:testing}

\begin{figure}
\includegraphics[width=0.99\columnwidth]{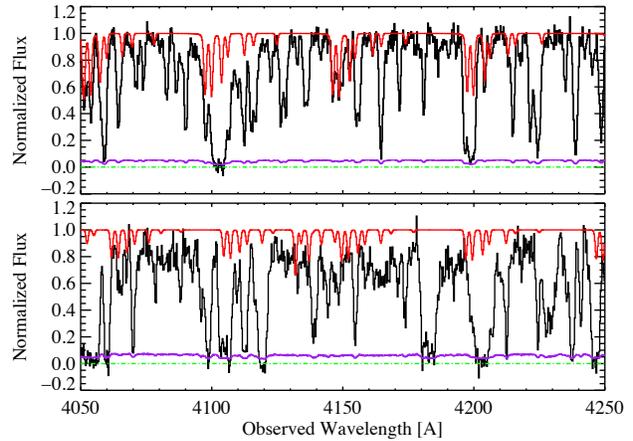}\\ 
\caption{ Example sections of spectra from a synthetic spectrum (top) compared with a MagE spectrum (bottom).  The synthetic spectrum was made with SNR = 19, a \dla\ at $z = 3$ and N(\htwo , J=1) = 10$^{18}$ cm$^{-2}$. The \htwo\ model is overplotted in red.  The MagE spectrum (bottom) is of J1246$+$1113 with a \dla\ at $z = 3.0971$, SNR$\sim$17 and N(\htwo , J=1) $\leq$ 10$^{17.1}$ cm$^{-2}$ (model overplotted in red).
}
\label{fig:synth}
\end{figure}

\begin{figure*}
\includegraphics[width=1.75\columnwidth]{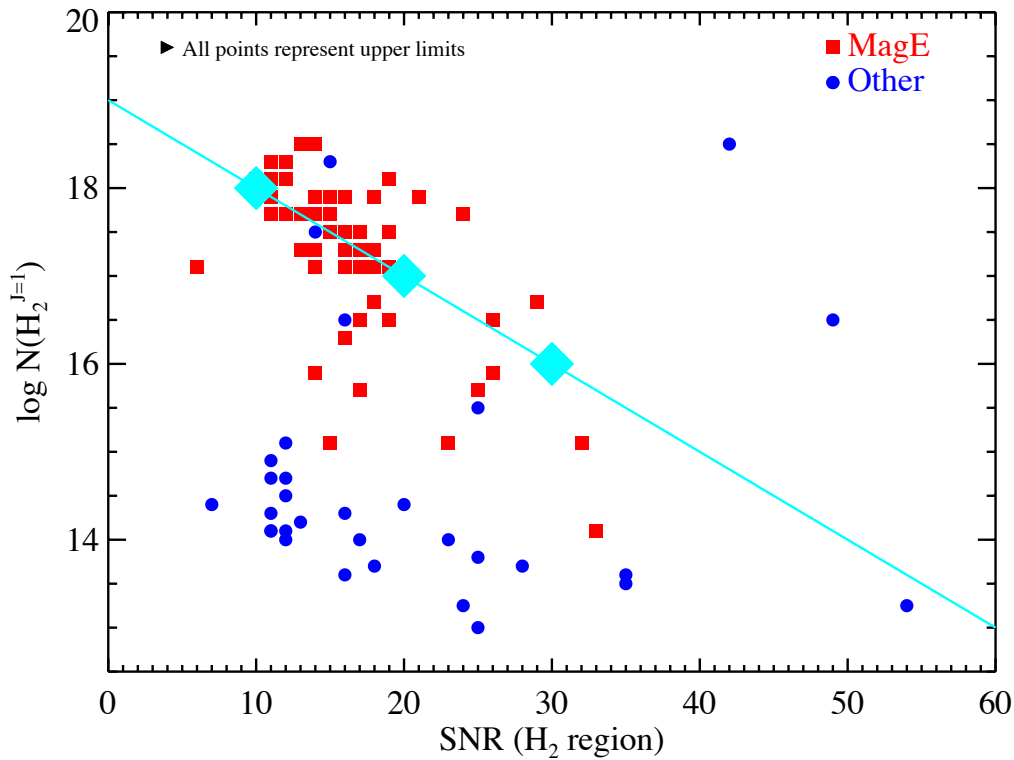}
\caption{1-$\sigma$ upper limits on the \htwo\ column density derived from the $J$=1 rotational level versus the SNR over the \htwo\ region of the spectrum for Magellan/MagE spectra (red squares) and all other data (blue circles).  The mean SNR$\sim$18.  Large cyan diamonds indicate the approximate log N(\htwo $^{J=1}$) that can be `recovered' by the upper limits code in synthetic spectra.  The cyan solid line indicates the analytical best-fit to these points as described in the text.
}
\label{fig:nh2vsnr}
\end{figure*}

In this section we perform two tests to determine the reliability of the upper limits routine, the first using synthetic spectra and the second using the UVES spectra of the N08 \htwo\ detections.  

\subsection{Analysis testing: recovery of a synthetic \htwo\ signal?}~\label{sec:sens}

In order to test the reliability of the upper limits algorithm, we analyzed synthetic spectra containing various amounts of molecular hydrogen and simulated\footnote{Lyman $\alpha$ forest simulated using code by ~\cite{liske08}.}  Lyman $\alpha$ forest lines.  To best match the data, the synthetic spectra were created with the resolution of the Magellan/MagE spectra, FWHM $\sim$ 71\kms , and a variety of signal-to-noise ratios (SNR).  Specifically, we created a synthetic  quasar spectrum containing a \dla\ at $z_{abs}=3.0$, the associated atomic hydrogen Lyman series lines,  and various amounts of molecular hydrogen.   Following the implementation of the upper limit code on the actual data, we assumed a model of temperature T=100 K for populating the $J$=0 and $J$=2 states, for a given $J$=1 column density.  We also included a simulated  Lyman $\alpha$ forest, in which the density of forest lines increases with redshift, in order to estimate the smallest amount of \htwo\ that the code was sensitive to picking out at a given resolution and SNR.  In performing this exercise we took the \htwo\ Doppler parameter to be $b$ = 2.0 km s$^{-1}$, a conservative choice given the above discussion of typical \htwo\ Doppler parameters. As we steadily decreased the input \htwo\ column density, we defined the minimum \htwo\ column density that could be recovered as the point at which the upper limit column density returned by the code was 0.1 dex above the input column density.  In other words, as the input column density was further decreased, the returned upper limit became more divergent from the input column density than 0.1 dex.  
 
\begin{figure*}
\includegraphics[width=1.75\columnwidth]{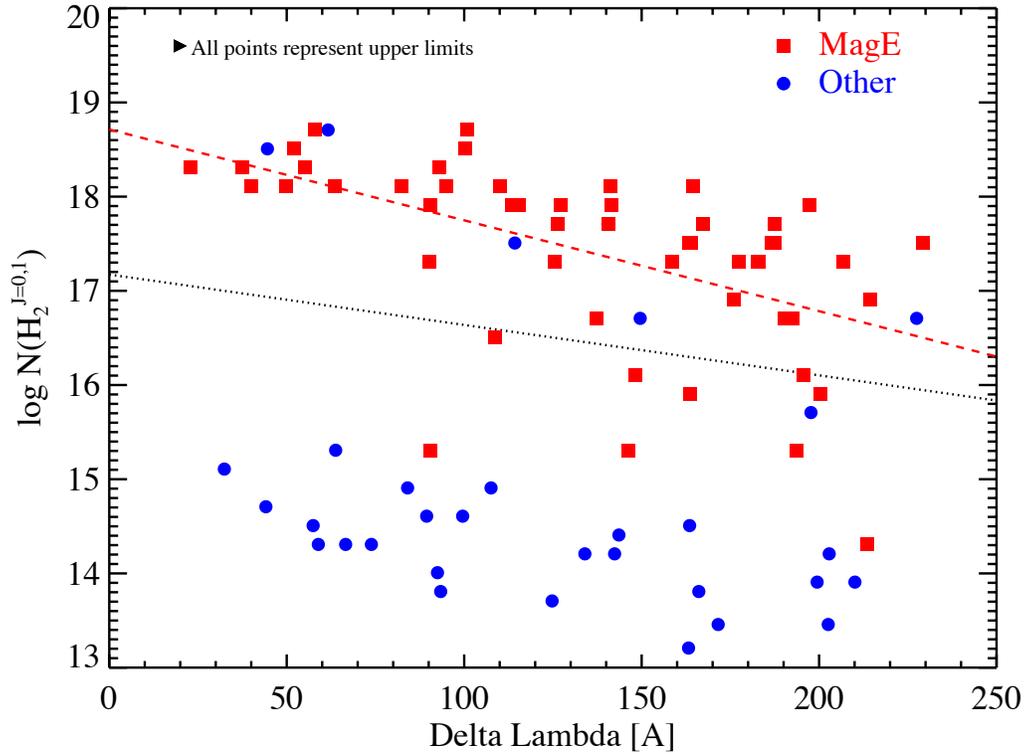}
\caption{Upper limits (1-$\sigma$) on the \htwo\ column density derived from the $J$=0 and 1 rotational levels, assuming $b$ = 2 \kms\ versus the $\Delta$$\lambda$ available to search for \htwo .  It is apparent that the larger $\Delta$$\lambda$ is, the stronger the log N(\htwo $^{J=0,1}$) upper limits become.  The red dashed line describes the best-fit power law fit to the Magellan/MagE (FWHM $\sim$ 71 \kms ) data \emph{only}, as described in the text, while the black dotted line is the best fit for all of the data plotted. 
}
\label{fig:nh2vdeltalam}
\end{figure*}

Naturally, the results of this test are dependent upon the SNR.  For SNR = 18\,per 20\,km\,s$^{-1}$ pixel, the mean of the sample, we find that for an input N(\htwo , $J$=1) = 10$^{17.0}$ $\cm{-2}$, the upper limit code returns N(\htwo , $J$=1) $<$ 10$^{17.1}$ $\cm{-2}$,   whereas for input N(\htwo , $J$=1) = 10$^{16.5}$ $\cm{-2}$ the upper limit code returns N(\htwo , $J$=1) $<$ 10$^{16.9}$ $\cm{-2}$.  Hence the  Lyman $\alpha$ forest confusion at this SNR (18) is enough that the upper limit returned is significantly greater (by 0.4 dex) than the input \htwo\ column density.  Therefore, we take N(\htwo , $J$=1) $=$ 10$^{17}$ $\cm{-2}$ as the approximate sensitivity limit for SNR $\sim$ 20 spectra.  For SNR = 10, we find that this value is N(\htwo , $J$=1) $\sim$ 10$^{18}$ $\cm{-2}$ and for SNR = 30, N(\htwo , $J$=1) $\sim$ 10$^{16}$ $\cm{-2}$.  While in reality each spectrum probes a unique line of sight through the  Lyman $\alpha$ forest, these values provide, on average, some indication of the level of sensitivity to detecting \htwo\ of the MagE spectra.  In Figure~\ref{fig:synth} we show an example section of synthetic spectrum along with a MagE spectrum for comparison.

To compare these levels with the upper limits determined from the data, we plot in Figure~\ref{fig:nh2vsnr} the upper limits on log N(\htwo , $J$=1) as determined for $b$ = 2.0 \kms\ from the upper limits code versus the SNR of the \htwo\ spectral region over which the upper limits were calculated.  In general, we limited our sample to spectra with SNR $\geq$ 10, however, we have included two \dlas\ with SNR $\sim$ 6 over the \htwo\ region because there was a significant number of \htwo\ lines available allowing for a relatively strong limit despite the reduced SNR.  As expected, it is clear that the SNR makes a large difference,  i.e. the low SNR spectra tend to have weaker, not as constraining limits -- note the cluster of points between SNR = 10 $-$ 20 and upper limits of N(\htwo ) = 10$^{18}$ $\cm{-2}$.  We have overplotted as cyan diamonds the approximate `sensitivity limits' as determined above.  Overplotted as a cyan line is the following simplified power law model, 

\begin{equation}
\mathrm{log N}(H_{2}) = a + b\ \mathrm{(SNR)}
\label{eqn:snr}  
\end{equation}      

\noindent where a = 19.0 and $b$ = $-$0.1.  All \dlas\ close to or above this line have been carefully inspected and denoted as either possible \htwo -bearing \dlas\ in need of followup high-resolution observations, or as possessing some trait, e.g. limited \htwo\ spectral coverage, that has limited the effectiveness of the upper limit code.   For example, \dla\ 1100$+$1122 contains an upper limit N(\htwo , $J$=1) $\leq$ 10$^{18.5}$ $\cm{-2}$ at SNR=42, but is also located at the relatively high redshift of $z_{abs}$ $\sim$ 3.75 where the forest lines are thicker, making it more difficult to rule out the presence of \htwo\ without a higher resolution spectrum. 
  
\begin{figure}
\includegraphics[width=0.99\columnwidth]{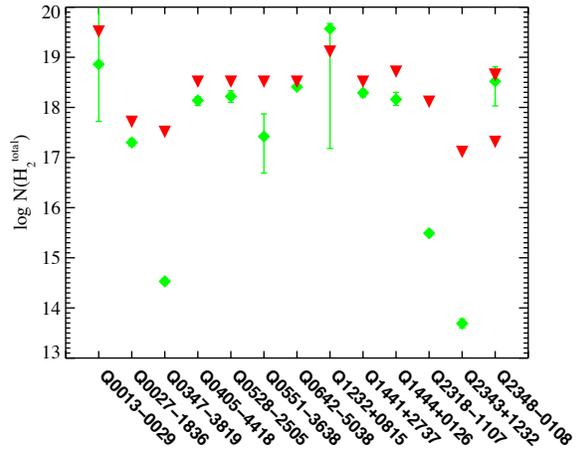}\\
\caption{A comparison of the \htwo\ column densities measured in the N08 UVES sample (FWHM$\sim$8 km s$^{-1}$, green diamonds with error bars) with the results of smoothing those same spectra to MagE resolution (FWHM$\sim$71 km s$^{-1}$) and using the upper limits algorithm to determine the allowed maximum amount of \htwo\ (red upside down triangles). It is clear that for most \dlas , the upper limit algorithm applied to the smoothed, lower resolution data provides upper limits on \htwo\ content consistent with that measured in the high resolution data.  Individual cases are discussed in the text.
}
\label{fig:smootheduves}
\end{figure}

Looking carefully at the 10 \dlas\ with N(\htwo ) upper limits greater than or equal to N(\htwo ) = 10$^{18}$  $\cm{-2}$ (assuming the conservative Doppler parameter, $b$ = 2 km s$^{-1}$) in Figure~\ref{fig:nh2vsnr}, we find that the majority of them have weak upper limits for reasons that may be completely unrelated to the potential \htwo\ content.  For example, 6 of the \dlas\ with upper limits N(\htwo ) $\geq$ 10$^{18}$ $\cm{-2}$, are likely due not to the presence of \htwo\ but rather to the fact that the spectral region available to search for \htwo , i.e. the spectral region having SNR $\geq$ 10, denoted under $\Delta$$\lambda$ in Table~\ref{tab:h2sample}, is less than $\approx$60 $\AA $.  This means that the upper limits were determined using only a few ($\lesssim$20) \htwo\ transitions.  Obviously, this scenario can result in non-constraining upper limits on the \htwo\ content that have little to do with an actual \htwo\ detection, merely because the probability of forest contamination influencing the result is increased.  
Indeed, a strong anti-correlation of N(\htwo ) upper limit value and spectral region searched, $\Delta$$\lambda$, can be seen in Figure~\ref{fig:nh2vdeltalam}.  A linear least-squares fit to the data,

\begin{equation}
\mathrm{log N}(H_{2}) = c + d\ \mathrm{(\Delta \lambda)}
\label{eqn:lam}  
\end{equation}    

\noindent provides c = 17.17 and d = $-$5.4$\times$10$^{-3}$ for a fit including all of the data, whereas c = 18.71 and d = $-$9.6$\times$10$^{-3}$ for a fit including \emph{only} the Magellan/MagE data.  We can combine equations~\ref{eqn:snr} and ~\ref{eqn:lam} into a single equation that provides approximate guidance on the required SNR and $\Delta$$\lambda$ necessary to achieve a particular N(\htwo ) sensitivity for spectra of Magellan/MagE or similar resolution (FWHM$\sim$71 km s$^{-1}$), 

\begin{equation}
\mathrm{log N}(H_{2}) = \frac{1}{2} (a + c + b\ \mathrm{(SNR)} + d\ \mathrm{(\Delta \lambda)})
\label{eqn:comb}  
\end{equation}    

\noindent where the constants are given above.

One of the 10 weak limit \dlas\ is \dla\ 1100$+$1122, discussed above, which is at a relatively high redshift, $z_{abs}$=3.75, and therefore suffers from thick forest confusion.  Finally, one more \dla\ is likely not a strong \htwo -bearer as it contains a very low metallicity, [M/H] $\sim$$-2.5$, and hence, is not likely to contain \htwo , considering the strong correlation between metallicity and \htwo\ content ~\citep{petitjean06}.  That leaves only 2 of the 10 \dlas , \dla\ 0127$+$1405 and \dla\ 1004$+$0018, to be possible \htwo -bearing \dlas\ at a similar level to that found by ~\cite{noter08}, assuming $b$ = 2 km s$^{-1}$. Indeed, these \dlas\ should be targeted for high-resolution follow-up spectroscopy in order to confirm or rule out the presence of \htwo .  

If we include the 2-3 weak limit \dlas\ that may contain significant \htwo , with the 1 confirmed \htwo\ detection, the \htwo\ covering factor of the Magellan sample increases to $\sim$4/86 = $\sim$5\,\%.   We note that, while it is not likely given the arguments presented above, even if all of the 10 weak limit \dlas\ in the Magellan sample actually contain \htwo , the \htwo\ covering factor, including the 1 confirmed \htwo\ detection, would be 11/86, or $\sim$13\,\%, a fraction still at odds with the 18\,\% expected from the N08 sample ~\citep{noter08}. 

\subsection{Analysis testing: recovery of the N08 \htwo\ detections?}~\label{sec:uvesrecovery}
In order to further test the validity of our upper limits algorithm and our ability to detect \htwo\ in medium-resolution spectra, we attempted to detect the N08 \htwo\ detections ~\citep{noter08} after smoothing the UVES spectra (FWHM$\sim$8 km s$^{-1}$) to the resolution of the MagE spectra (FWHM$\sim$71 km s$^{-1}$) used in this survey.  In order to reproduce, as closely as possible, the process we followed with the MagE spectra, we first convolved the un-normalized UVES spectra with a Gaussian kernel and rebinned it to produce a spectrum of MagE resolution (FWHM$\sim$71 km s$^{-1}$). We then fit the quasar continuum using the same process as that used for the MagE spectra (see \S~\ref{sec:continuum}) and added noise such that the SNR of the smoothed spectrum matched that of our typical MagE spectra, SNR$\sim$20 pixel$^{-1}$.  Finally, we passed these smoothed spectra through the same upper limits algorithm to search for \htwo , assuming a Doppler parameter $b = 2$ km s$^{-1}$ and T = 100 K for populating the J levels, as described in \S~\ref{sec:analysis}.  We present the results of this search in Figure~\ref{fig:smootheduves}, where we plot the log of the \htwo\ column density upper limits derived from the smoothed UVES spectra as red upside down triangles for each of the \htwo\ detections in the N08 sample.  Overplotted as green diamonds with error bars are the UVES \htwo\ measurements from N08. 

It is clear that in almost every case, the upper limits algorithm returns an \htwo\ upper limit from the smoothed UVES spectrum that is consistent with the N(\htwo ) measured in the original high-resolution spectrum.  In fact, this agreement is remarkable given the lower resolution, (typically) decreased SNR, and simplifying assumptions made in order to reproduce the MagE process, such as a single velocity component with Doppler parameter $b = 2$ km s$^{-1}$ and T = 100 K for populating the J levels.  In all cases we report the total amount of \htwo\ assuming T=100 K for populating the $J$ = 0 and 2 states.

We now discuss a few individual cases involving the \dlas\ towards: 

$\bullet$ Q0347$-$3819, Q2318$-$1107 and Q2343$+$1232: The \htwo\ column density in these three \dlas\ is small enough that we would {\it not} have expected to detect it in our MagE-resolution survey.  At MagE resolution and SNR$\sim$20, as shown in Figure~\ref{fig:nh2vsnr}, we are sensitive down to N(\htwo ) $\sim$ 10$^{17}$ cm$^{-2}$, and therefore we would not have expected to detect these N(\htwo ) $<$ 10$^{16}$ cm$^{-2}$ systems.  

$\bullet$ Q0551$-$3638:  The SNR of the original UVES spectrum in the region of \htwo\ transitions is already relatively low at SNR $\lesssim$ 8. This fact, in combination with a relatively small $\Delta \lambda$ for detecting \htwo\ ($\Delta \lambda$ $\sim$ 50 \AA ) resulted in an upper limit in the smoothed spectrum of N(\htwo ) $\sim$ 10$^{18.5}$ cm$^{-2}$, not quite sensitive enough to detect the actual N(\htwo) = 10$^{17.4}$ cm$^{-2}$ measured in the high resolution UVES spectrum.

$\bullet$ Q1232$+$0815:  For this \dla\ we show a range of error bars to include the range of published \htwo\ column densities, from N(\htwo ) = 10$^{17.18}$ cm$^{-2}$ ~\citep{sri00}, to N(\htwo ) = 10$^{19.57}$ cm$^{-2}$ ~\citep{balashev11}.  Taken at face value the UVES spectrum appears to have zero-level problems that were later interpreted by ~\cite{balashev11} to be the result of partial covering of the broad line region by the \htwo\ cloud.  Despite these issues, the naive fitting of the degraded resolution spectrum still provides a reasonable upper limit of N(\htwo ) = 10$^{19.11}$ cm$^{-2}$.

$\bullet$ Q2348$-$0108: In this case we have placed two upper limits to the \htwo\ column density, the lowest of which (N(\htwo ) $<$ 10$^{17.31}$ cm$^{-2}$) was derived in the simplistic way using a single velocity component with the previously stated assumptions.  However, in this case, the \htwo\ absorption is spread over several velocity components, two of which contain the bulk of the \htwo\ and are separated by $\sim$26 km s$^{-1}$.  If we instead determine the upper limits at the redshifts of the two strongest \htwo\ components and sum them, we find N(\htwo ) $<$ 10$^{18.65}$ cm$^{-2}$, consistent with the total N(\htwo ) = 10$^{18.52}$ cm$^{-2}$, measured in the UVES spectrum.

This last case, of the \dla\ towards Q2348$-$0108, raises the issue of how potential redshift offsets of \htwo\ relative to the low-ions that determine $z_{abs}$ may affect our upper limits in the Magellan survey and we address this issue in \S~\ref{sec:redshiftoffsets}.

\begin{figure*}
\includegraphics[width=1.75\columnwidth]{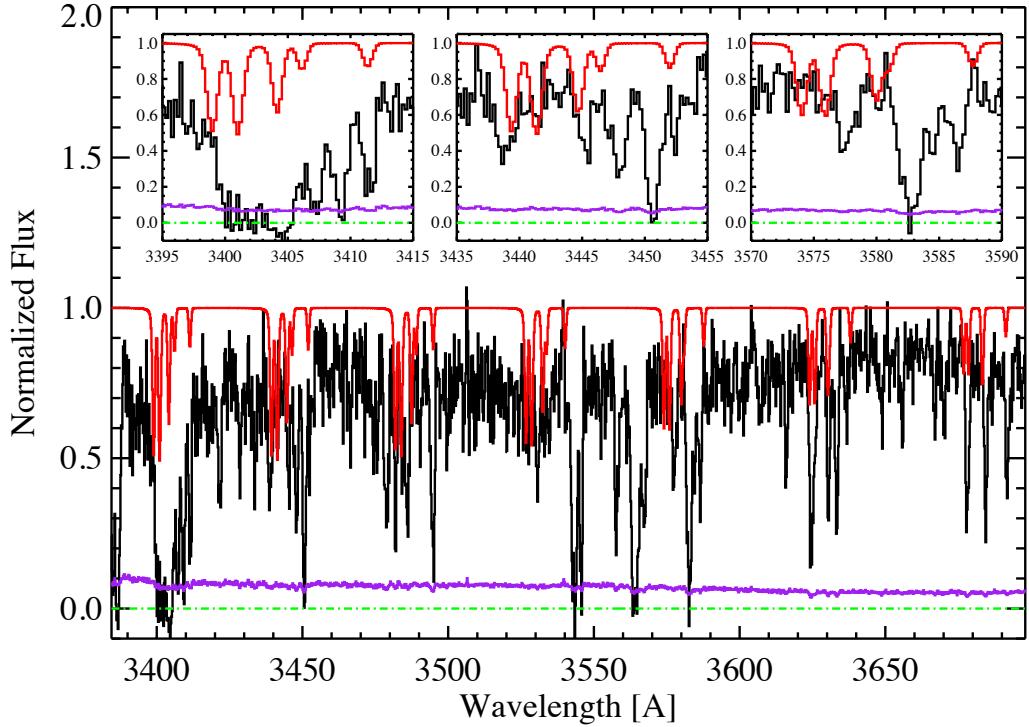}
\caption{The Lyman and Werner band \htwo\ transition region in MagE spectrum of the \dla\ towards J0234$-$0751. Overplotted in red is the model fit where N(\htwo\ $^{J=1}$) = 18.1 cm$^{-2}$ and $b = 2$ km s$^{-1}$.  All \htwo\ transitions used in the fit (here, 35) are shown, while the three insets above provide a closer look at some of them.  The top left inset contains no `constrained' transitions, while the middle inset contains three `constrained' transitions and the right inset contains four. 
}
\label{fig:constrained}
\end{figure*}

\subsubsection{Estimating the likelihood that a weak \htwo\ upper limit may be a detection}

When the upper limits algorithm returns a `weak' upper limit, defined here to be a limit with N(\htwo ) $\geq\ \sim$10$^{18}$ cm$^{-2}$, it is the result of three possible physical scenarios: 1) the presence of strong (N(\htwo)$>$10$^{18}$cm$^{-2}$) \htwo\ absorption, made obvious by several unblended \htwo\ transitions, 2) the presence of strong \htwo\ absorption that is not obvious, perhaps constrained only by a single transition, or 3) the absence of strong \htwo\ absorption in conjunction with any one of, or a combination of, the following: small wavelength coverage in the observed spectral range of the \htwo\ transitions, low spectral SNR, and/or unlucky blending with the Lyman $\alpha $ forest.  While our visual inspection as described in Section~\ref{sec:analysis} rules out scenario 1, we now attempt to develop a method of differentiating between scenarios 2 and 3, i.e. to determine whether or not strong, non-obvious \htwo\ may be present.

\input{MNRAS_likelihood.tex}

With the goal of differentiating these two cases and determining which `weak' upper limits are likely to be due to the presence of \htwo\ rather than simply the unfortunate circumstance of the spectrum, we have developed a likelihood estimator for the presence of \htwo .  In order to accomplish this we overplot the model \htwo\ fit determined by the upper limit routine, in this case assuming $b = 2$ km s$^{-1}$, and determine the number of \htwo\ transitions that were used to constrain the upper limit.  A constrained transition is defined as one in which the model fit either equals or violates the data, see Figure~\ref{fig:constrained} for an example.  In the weak limit regime, the more transitions that were used to constrain the upper limit, the more likely the upper limit is created by the presence of \htwo .  In other words, if many \htwo\ transitions constrain the weak limit fit and none rule out strong \htwo , it is more likely that \htwo\ is present at the detected level.  On the other hand, if only one or two transitions constrain the fit, the weak upper limit may simply be a reflection of small wavelength coverage, low SNR, or significant blending with the Lyman $\alpha$ forest.

\begin{figure}
\includegraphics[width=0.99\columnwidth]{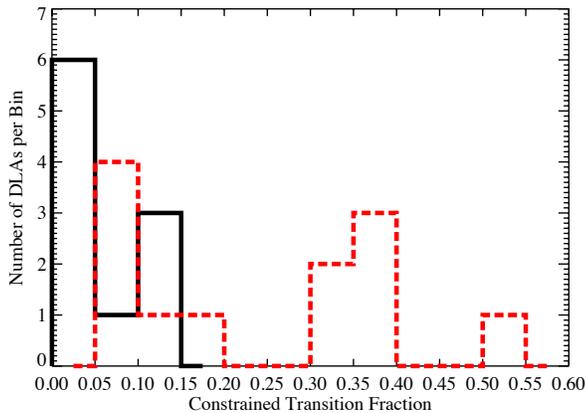}\\
\caption{The constrained transition fraction, $C = \frac{constrained}{total}$, of the weak N(\htwo ) upper limits for the smoothed UVES detections (red dashed histogram) and the Magellan sample (black histogram). It is clear that the likelihood of the Magellan sample upper limits to be caused by the presence of \htwo , rather than some other feature, is relatively low compared with that of the UVES sample when smoothed to MagE resolution.  
}
\label{fig:likelihood}
\end{figure}

We define the constrained transition fraction, `C,'  of real \htwo\ content as $C = \frac{constrained}{total}$,  
where `constrained' is the number of \htwo\ transitions used to constrain the fit and `total' is the total number of transitions possible, defined as all transitions of \htwo\ $J = 0, 1, 2$, and equal to 134. The results, for the case of $b = 2$ km s$^{-1}$, are summarized in Table~\ref{tab:likelihood} and plotted as a histogram in Figure~\ref{fig:likelihood}.  It is clear from Figure~\ref{fig:likelihood} that the likelihood of the smoothed UVES upper limits, determined in ~\S~\ref{sec:uvesrecovery} and shown in red, to contain \htwo\ are generally much higher than than those of the weak limit Magellan sample, which never reach a constrained transition fraction above $\sim$0.15. 

If we assume that the 3 \dlas\ of the Magellan sample with higher constrained transition fractions might actually contain \htwo\ we find that, including the 1 already detected \htwo\ absorber, the rate of detection of the Magellan sample would be 4/86 = $\sim$5\,\%, smaller than that of N08 (10 $\pm$ 4\,\%) at $\sim$89\,\% confidence.

\begin{figure}
\includegraphics[width=0.99\columnwidth]{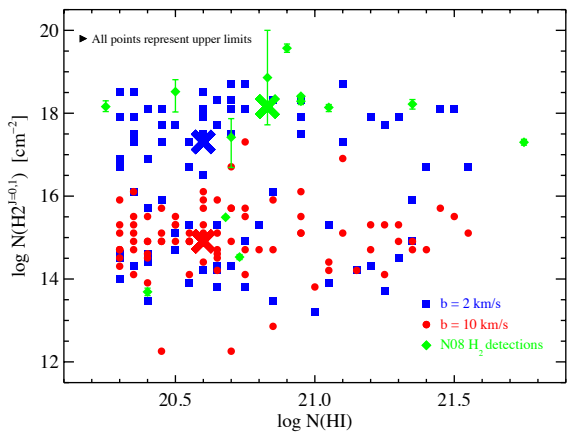}\\
\caption{\htwo\ upper limits versus the log \nhi .  It is clear that there are no trends with \nhi\ value. 
}
\label{fig:nh2vnhi}
\end{figure}

\section{Impact of Potential systematic Errors}\label{sec:problems}
In this section we discuss several potential issues that may have affected the results of the Magellan survey for \htwo . We first discuss the issue of potential redshift offsets between the low-ion absorption lines and possible \htwo\ absorption.  We then analyze both the Magellan and the N08 samples for any trends of \htwo\ detections (or limits) with redshift, metallicity or \nhi\ distribution as well as with quasar or spectral properties that might be indicative of biases.  Finally, we discuss the implications of potential continuum estimation errors.

\subsection{Impact of potential redshift offsets}\label{sec:redshiftoffsets}

As seen in the \dla\ towards Q2348$-$0108 ~\citep{noter07} and discussed in a previous section (\S~\ref{sec:uvesrecovery}), the \htwo\ velocity component structure may be complex and consist of multiple velocity components and/or be different than that of the low-ions from which $z_{abs}$ is determined.  Either of these situations may be problematic for our method of determining the \htwo\ upper limits in the MagE spectra using the redshift of the highest optical depth low-ion velocity component.  

However, we took special care to search for \htwo\ at multiple possible redshifts in \dlas\ with large $\delv$ and/or multiple strong low-ion velocity components.  We then report the least constraining \htwo\ upper limit.  In this way, we account for the possibility that the bulk of the \htwo\ may be offset from the main low-ion component.  

Finally, we note that this problem is likely not to have a large impact on our survey.  A review of the literature shows that in $\sim$76\,\% (13 out of 17) of \dlas\ containing \htwo , the \htwo\ velocity component containing the bulk of the \htwo\ coincides with the low-ion velocity component of highest optical depth (that which would be used to determine $z_{abs}$) to within a few ($\sim$2) km s$^{-1}$ or less. 
Therefore, by initially searching for \htwo\ at the redshift of the strongest low-ion component, we are simply pre-selecting the most likely location of \htwo .

\subsection{Trends with \dla , quasar or spectral properties?}
In order to rule out the presence of trends in upper limits with \dla , quasar or spectral properties, we investigated the \htwo\ column density upper limits as a function of metallicity, \nhi , $z_{abs}$, and $I$-band magnitude of the background quasar. As with the metallicity (Figure~\ref{fig:nh2vm}), there are no apparent trends in the upper limits measurements with \nhi , $z_{abs}$ or $I$-band magnitude of the background quasar.  In Figure~\ref{fig:nh2vnhi} we plot the N(\htwo ) upper limits versus the log \nhi\ of the \dla .  We find no evidence for any trends of N(\htwo ) with \nhi\ value.  There is also no trend for larger \nhi\ values to have weaker N(\htwo ) upper limits.  This result lends confidence that the upper limits measurements are secure and not biased by some unrelated, external factor.

\subsection{Comparing the \nhi , metallicity and redshift distributions of the Magellan and N08 samples}\label{sec:distributions}

\begin{figure}
\includegraphics[width=0.99\columnwidth]{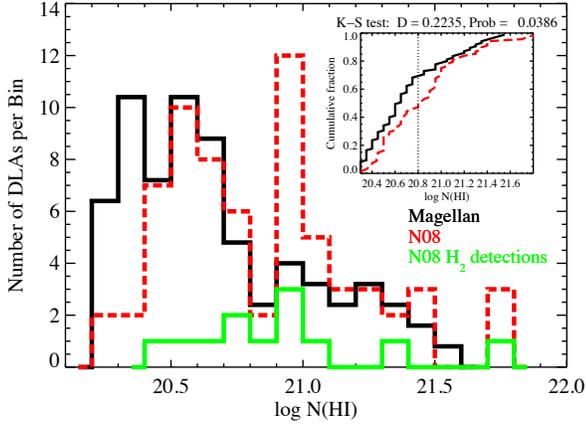}\\
\caption{\nhi\ histogram comparing the N08 ~\citep{noter08} distribution in red (dashed) with the scaled Magellan sample in black.  A K-S test indicates the probability they are drawn from the same parent population is P$_{KS} \sim$ 0.039.  Overplotted in green is the distribution of the \htwo\ detections from the N08 sample.  
}
\label{fig:nhidist}
\end{figure}

\begin{figure}
\includegraphics[width=0.99\columnwidth]{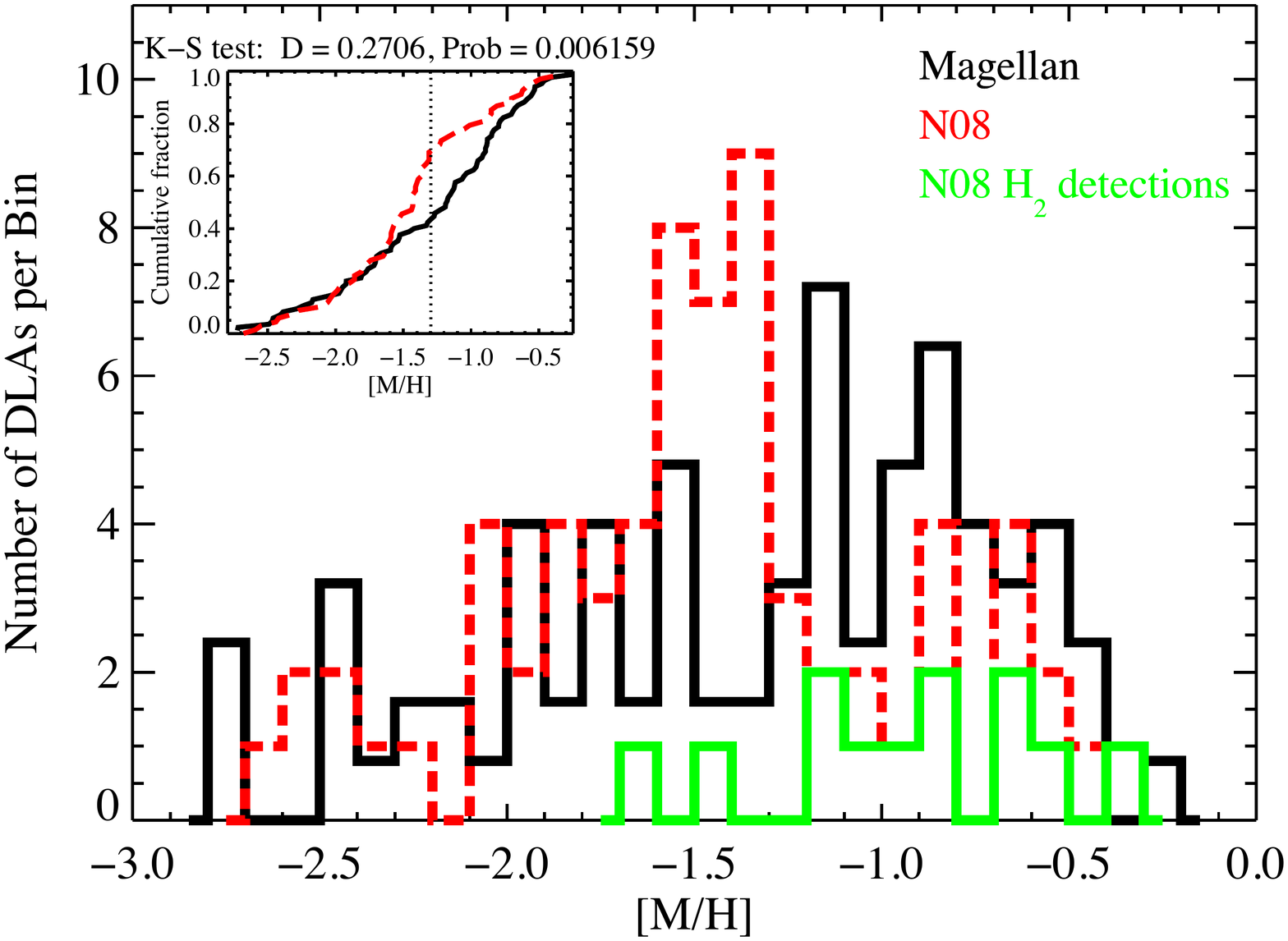}\\
\caption{Metallicity histogram comparing the N08 ~\citep{noter08} distribution in red (dashed) with the scaled Magellan sample in black.  A K-S test indicates the probability they are drawn from the same parent population is P$_{KS} \sim$ 0.006.  Overplotted in green is the distribution of the \htwo\ detections from the N08 sample.
}
\label{fig:metdist}
\end{figure}

\begin{figure}
\includegraphics[width=0.99\columnwidth]{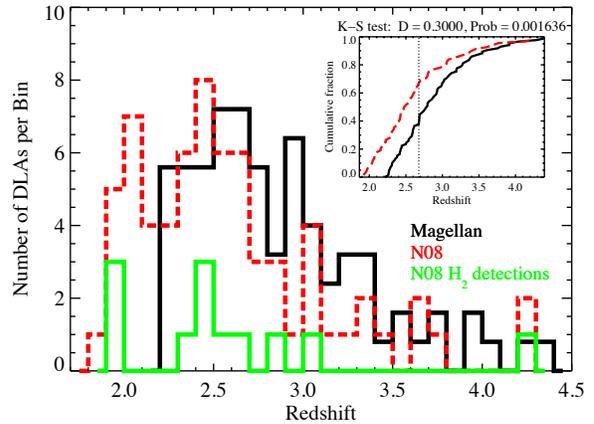}\\
\caption{Redshift histogram comparing the N08 ~\citep{noter08} distribution in red (dashed) with the scaled Magellan sample in black.  A K-S test indicates the probability they are drawn from the same parent population is P$_{KS} \sim$ 0.002.  Overplotted in green is the distribution of the \htwo\ detections from the N08 sample.
}
\label{fig:zabsdist}
\end{figure}

In Figures~\ref{fig:nhidist},~\ref{fig:metdist} and~\ref{fig:zabsdist} we compare the distributions of \nhi , metallicity and redshift,  respectively, of the Magellan sample with those of the 68 bonafide \dlas\ of the N08 sample.  The Magellan sample, in black, has been normalized such that the area under the histogram equals the area under the N08 sample histogram, in red.  To compare with the distribution of the \htwo\ detections, we overplot the \htwo\ detections of the N08 sample in green.

 It is clear that in all cases the distributions are quite different from each other and have a low K-S probability, $< 4\,\%$, of being drawn from the same parent population.  As seen in Figure~\ref{fig:nhidist} and already discussed in ~\S~\ref{introduction} and in ~\cite{noter08},  the \nhi\ distribution of the N08 sample is biased towards large \nhi\ while simultaneously under-sampling the low-\nhi\ regime.  It is interesting to note that the \htwo\ detections, in green, trace the high \nhi\ overdensity of the parent N08 sample. 
 
In Figure~\ref{fig:metdist} we compare the metallicity distributions of the two samples and see that the \htwo\ detections are preferentially found in the high metallicity regime.  It is interesting to note that the peak of the Magellan sample metallicity distribution is actually higher than that of the N08 sample, indicating that the Magellan sample is slightly more biased towards higher metallicity systems and, if anything, should have found more \htwo , according to the ~\cite{petitjean06} metallicity-\htwo\ correlation.  

Finally, in Figure~\ref{fig:zabsdist} we examine the redshift distributions of the two samples.  It is clear that these distributions are quite different, with the most obvious difference being that the Magellan sample, by definition, stops at $z = 2.2$, whereas the N08 sample contains \dlas\ with redshifts as low as $z = 1.864$. If we consider the N08 sample only in the range of the Magellan sample, i.e. $z \ge\ $ 2.2, we find a K-S probability P$_{K-S}$ = 0.17 that the samples share the same parent population.  Hence, when compared over a similar redshift range, the redshift distributions of the Magellan and N08 samples are not altogether different.   While it is possible that the \htwo\ detection rate of the N08 sample is larger than that of the Magellan sample presented here because of an increase of \dla\ \htwo\ content below $z = 2.2$, it seems unlikely that any significant increase could take place over the relatively short time span of $\approx$0.5 Gyr. 

While the differences mentioned above may help to explain the difference in \htwo\ content found by each sample, the uniformly-selected Magellan sample remains the only one to be created {\it a priori} in a way that was as blind and as unbiased as possible.

\subsection{Implications of continuum errors}~\label{sec:continuum}
One of the more difficult problems in working with quasar absorption lines is the determination of the original, unabsorbed quasar continuum level.  This issue is particularly pernicious in the region of the  Lyman $\alpha$ forest, where the forest absorption lines can be thick and blended, leading to an overall apparent lowering of the true quasar continuum level.  We first outline our method of continuum fitting the quasar spectra and then conclude with an analysis of errors due to potential errors in continuum fitting.  

Continuum fitting of the reduced quasar spectra was done using the XIDL\footnote{http://www.ucolick.org/$\sim$xavier/HIRedux/index.html} command x\_continuum, which allows for an interactive cubic spline fit through the data points.  Using the interactive graphical user interface we select sections of spectra to fit and a higher order spline fit is made through the data.  In this way, one can avoid spectral regions affected by absorption lines by fitting only the quasar continuum on either side of the absorption features.  Redward of the quasar  Lyman $\alpha$ emission peak, this procedure fits the quasar continuum well, with little need for additional by-hand adjustment.  However, blueward of the quasar Lyman $\alpha$ emission peak, the prevalence of the Lyman $\alpha$ forest absorption makes this procedure difficult as the unabsorbed regions of quasar spectra can be sparse and difficult to identify.  In this region, we followed the typically applied procedure of fitting the continuum by identifying the highest flux peaks (transmission peaks) in the spectra and fitting the continuum through these peaks.  We typically placed additional spline points by hand to produce the most featureless continuum consistent with the identified transmission peaks. Given that transmission peaks may suffer from non-obvious absorption, this procedure may underestimate the true quasar continuum, especially in cases where the forest contamination is heavy.

To investigate the possible implications of errors in the quasar continuum placement in this study, we determined the effects on the resultant upper limits when the quasar continuum is adjusted by $\pm$10\,\% and $\pm$20\,\%.  To get an idea of the overall effects of continuum changes, we highlight here the results of performing this test on a random sub-sample of \dlas\ with a representative range of upper limits spanning both the `strong' and `weak' limits regime, where `strong' refers to an upper limit that excludes the presence of significant \htwo , i.e. N(\htwo ) $\lesssim$ 10$^{15}$ cm$^{-2}$), and `weak' refers to upper limits with larger column density values such as N(\htwo ) $\sim$ 10$^{18}$ cm$^{-2}$. Note, in all cases we assume $b$ = 2 km s$^{-1}$.

\begin{figure*}
\includegraphics[width=1.75\columnwidth]{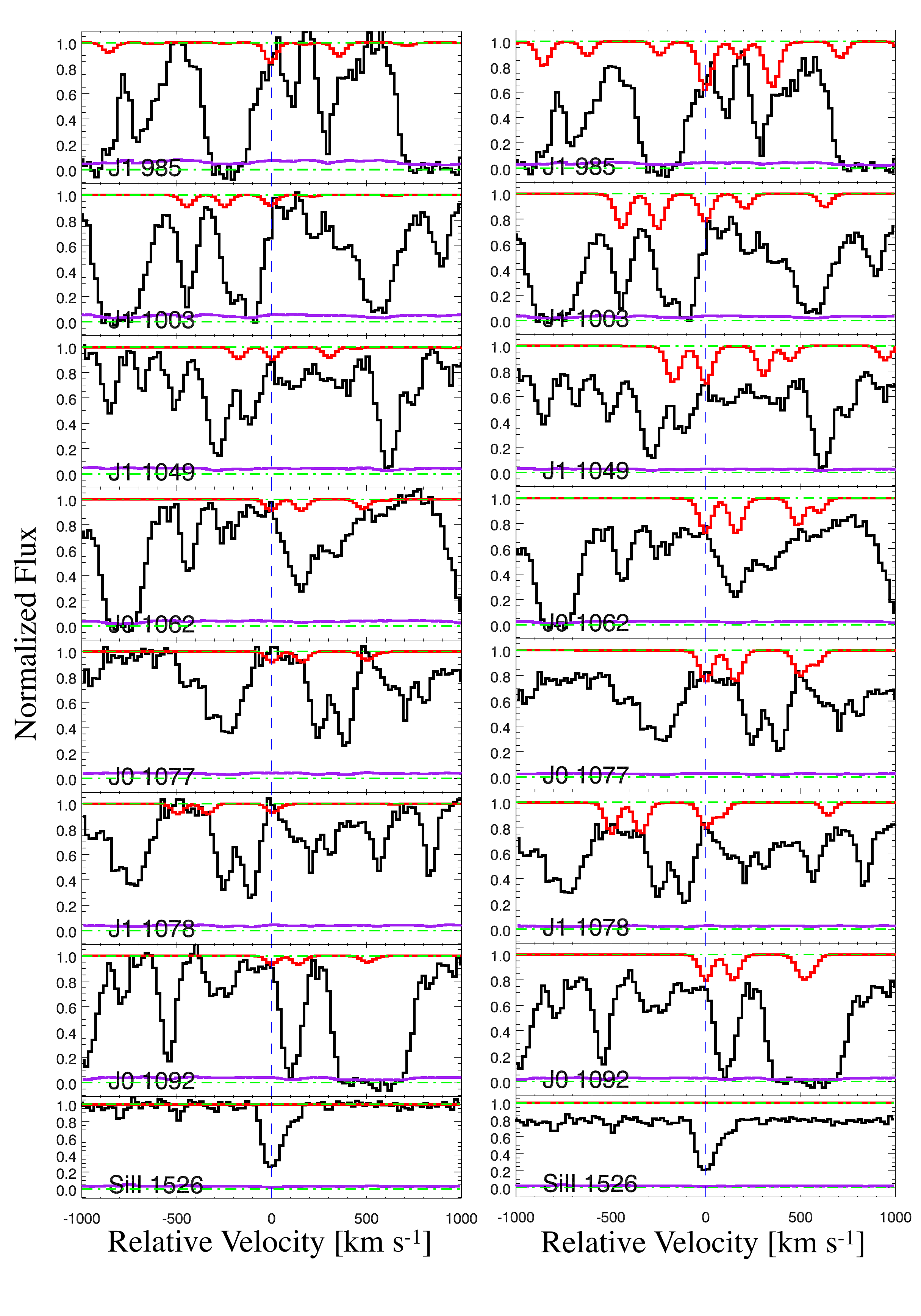}
\caption{An example of a strong \htwo\ upper limit and the effects of continuum changes on the resultant upper limits.  Plotted in black is a selection of \htwo\ $J$=0 and $J$=1 transitions of the $z_{abs}$=2.8691 \dla\ towards J2238$-$0921.  The left column contains the original continuum fit, while the right column contains a continuum fit that is 20\,\% higher.  Overplotted in red is an \htwo\ model assuming the measured N(\htwo ) upper limits for $b = $ 2 km s$^{-1}$, N(\htwo ) = 10$^{15.1}$ $\cm{-2}$ (left column) and N(\htwo ) = 10$^{17.5}$ $\cm{-2}$(right column).  Overplotted in purple is the 1-$\sigma$ error array.  The bottom panel shows the Si II $\lambda$1526 low-ion metal transition for reference. 
}
\label{fig:J2238}
\end{figure*}

\begin{figure*}
\includegraphics[width=1.75\columnwidth]{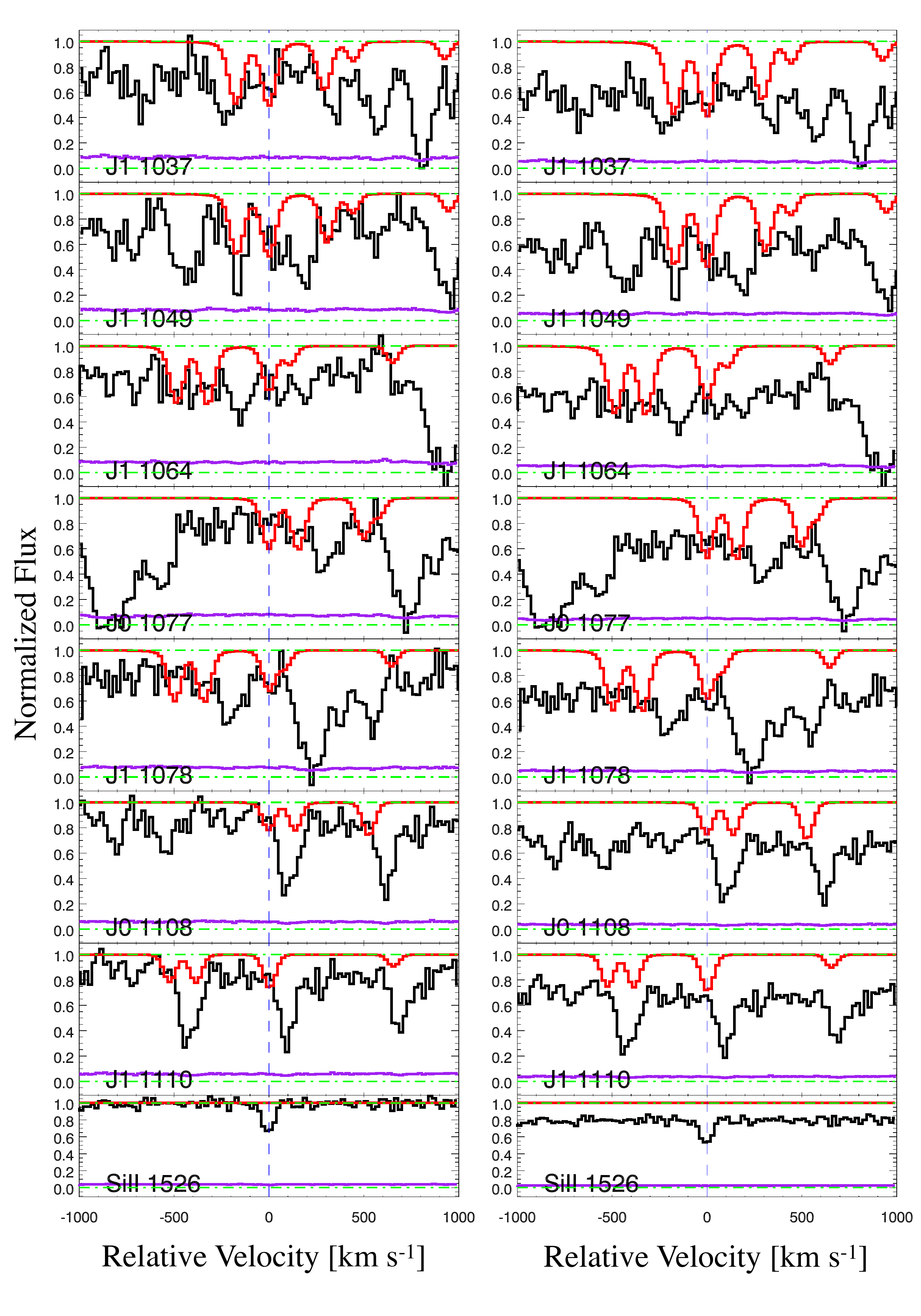}
\caption{An example of a weak \htwo\ upper limit and the effects of continuum changes on the resultant upper limits.  Plotted in black is a selection of \htwo\ $J$=0 and $J$=1 transitions of the $z_{abs}$ = 2.3181 \dla\ towards J0234$-$0751.  The left column contains the original continuum fit, while the right column contains a continuum fit that is 20\,\% higher.  Overplotted in red is the \htwo\ model fit assuming the measured N(\htwo ) upper limits, N(\htwo ) = 10$^{18.1}$ $\cm{-2}$ for $b = $ 2.0 \kms\ (left column) and N(\htwo ) = 10$^{18.3}$ $\cm{-2}$ for $b = $ 2.0 \kms\ (right column).  Overplotted in purple is the 1-$\sigma$ error array. The bottom panel shows the Si II $\lambda$1526 low-ion metal transition for reference.
}
\label{fig:J0234}
\end{figure*}

\begin{enumerate}

\item{For the $z_{abs}$=4.0209 \dla\ of $J$=1257$-$0111, where N(\htwo ) $\leq$ 10$^{14.1}$ $\cm{-2}$, raising the continuum by 10\,\% (20\,\%) produces the upper limit N(\htwo ) $\leq$ 10$^{14.1}$(10$^{15.9}$) $\cm{-2}$, while lowering the continuum by 10\,\% (20\,\%) returns N(\htwo ) $\leq$ 10$^{12.0}$ (10$^{12.0}$) $\cm{-2}$ (the lowest column density tested).}

\item{For the $z_{abs}$=2.8691 \dla\ of J2238$-$0921, where N(\htwo ) $\leq$ 10$^{15.1}$ $\cm{-2}$, raising the continuum by 10\,\% (20\,\%) produces the upper limit N(\htwo ) $\leq$ 10$^{16.5}$ (10$^{17.5}$) $\cm{-2}$, while lowering the continuum by 10\,\% (20\,\%) returns N(\htwo ) $\leq$ 10$^{12.0}$ (10$^{12.0}$) $\cm{-2}$ (the lowest column density tested).}

\item{For the $z_{abs}$=2.3181 \dla\ of J0234$-$0751 where N(\htwo ) $\leq$ 10$^{18.1}$ $\cm{-2}$, raising the continuum by 10\,\% (20\,\%) produces the upper limit N(\htwo ) $\leq$ 10$^{18.1}$ (10$^{18.3}$) $\cm{-2}$, while lowering the continuum by 10\,\% (20\,\%) returns N(\htwo ) $\leq$ 10$^{18.1}$ (10$^{17.7}$) $\cm{-2}$.}

\item{For the $z_{abs}$=2.540 \dla\ of $J$=1004$+$0018, where N(\htwo ) $\leq$ 10$^{18.5}$ $\cm{-2}$, raising the continuum by 10\,\% (20\,\%) produces the upper limit N(\htwo ) $\leq$ 10$^{18.5}$ (10$^{18.5}$) $\cm{-2}$, while lowering the continuum by 10\,\% (20\,\%) returns N(\htwo ) $\leq$ 10$^{18.3}$ (10$^{18.3}$) $\cm{-2}$.}

\end{enumerate}

As expected, the results can vary widely from system to system, depending on the nature of each spectrum and its  Lyman $\alpha$ forest contamination.  A careful inspection of these systems reveals that while SNR, $\Delta$$\lambda$ and the distribution of  Lyman $\alpha$ forest absorption all contribute, another factor in determining the effect on N(\htwo ) upper limits from raising the continuum level seems to be the number of pixels used to constrain the upper limit.  For example, if a particular upper limit is constrained by only a few pixels (say, a single \htwo\ transition), then raising the continuum level will likely result in an increase in the upper limit (depending on SNR) because it is still this single transition that constrains the fit. On the other hand, if an \htwo\ upper limit is constrained by many pixels spanning several different \htwo\ transitions (particularly in combination with a low SNR), then changes in the continuum level are not as likely to create large changes in the \htwo\ upper limit because many transitions still work to constrain the upper limit.

The $z_{abs}$ = 2.8691 \dla\ towards J2238$-$0921, case 2 above, is an example of the first scenario in which few transitions/pixels constrain the upper limit in combination with a relatively high SNR (SNR$\sim$23).  In Figure~\ref{fig:J2238} we plot all of the \htwo\ transitions associated with relatively unabsorbed continuum for the original continuum fit (left column) and for a continuum that is raised by 20\,\% (right column).  Overplotted in red are model \htwo\ spectra containing the \htwo\ column density upper limit value assuming $b$ = 2 km s$^{-1}$, N(\htwo ) = 10$^{15.1}$ cm$^{-2}$ (left) and N(\htwo ) = 10$^{17.5}$ cm$^{-2}$ (right).  It is clear that the high SNR in combination with the small number of pixels that constrain the upper limit have caused the (albeit large) 20\,\% change in continuum level to create a relatively large increase in the N(\htwo ) upper limit (2.4 dex).
On the other hand, for the $z_{abs}$=2.3181 \dla\ of J0234$-$0751 (case 3), the SNR is relatively low (SNR$\sim$12) and the number of transitions/pixels constraining the upper limit, albeit weakly, is larger.  Therefore, increasing the continuum level by 20\,\% (seen in the right column of Figure~\ref{fig:J0234}) produces little change (0.2 dex) in the already weak N(\htwo ) upper limit.  

The previous examples highlight how changes in the quasar continuum level can affect (or not) the measured N(\htwo ) upper limits.  While it is difficult to estimate the overall effects of potential underestimation of the true quasar continuum on the derived N(\htwo ) upper limits, we do make note of two important points: 1) Underestimation of the quasar continuum by 20\,\% is unlikely.  Typically, quoted errors on quasar continuum placement for spectra of this resolution and SNR are $\lesssim$10\,\% (e.g. ~\cite{faucher08}), and 2) Changes in the measured N(\htwo ) upper limits with 10\,\% continuum increase tend to be relatively small and generally not enough to move a system from the strong limit regime to the weak limit regime.  We therefore conclude that while continuum placement errors may have some effect on our reported N(\htwo ) upper limits, it is likely to be small.

\section{Summary}~\label{sec:summary}

We have presented the results from the first uniformly selected, blind survey for molecular hydrogen in \dlas\ with moderate-to-high resolution spectra.  The major result of this survey is that we did not unambiguously detect any new \htwo\ absorbers in our sample of 86 \dlas .  Given the 1 sample \dla\ already known to contain \htwo , this resulted in an \htwo\ detection rate of 1 out of 86, or $\approx$1\,\%, assuming that all of the upper limits are non-detections.  This detection rate of 1\,\% is smaller (at 99.82\,\% confidence) than the previously detected rate of strong \htwo\ systems (N(\htwo ) $\geq$ 10$^{18}$cm$^{-2}$), $10.3^{+3.1}_{-4.6}$\,\% in the N08 sample ~\citep{noter08}.   Given that the Magellan sample is relatively unbiased and that we are confident in our completeness at the N(\htwo ) $\geq$ 10$^{18}$cm$^{-2}$ detection threshold, we show that the covering factor of strong \htwo\ systems (N(H$_2$)$>10^{17.5}$\,cm$^{-2}$) in \dlas\ is likely to be $\sim$1\,\% (with a 2$\sigma$ upper limit of 6\%).

Without obvious \htwo\ detections (excepting the previously detected z$_{abs}$=2.4260 \dla\ in the line of sight towards SDSS235057.87$-$005209.9 ~\citep{petitjean06}), we endeavored to estimate the total amount of \htwo\ that \emph{could} be contained in our sample by measuring the upper limits on \htwo\ column density for each \dla .  The derived N(\htwo ) limit is, of course, highly dependent on the assumed Doppler parameter, $b$.  Lacking knowledge of the true Doppler parameter, we determined upper limits for a range of possible Doppler parameters, $b$ = 1 $-$ 10 km s$^{-1}$.  With the caveat that the Magellan sample contains a mix of resolutions (from FWHM$\sim$71 km s$^{-1}$ to $\sim$8 km s$^{-1}$), even for the conservative Doppler parameter of $b$ = 2 km s$^{-1}$, the median value of the N(\htwo ) upper limits is 0.85 dex less than the median value of the detections of the N08 sample.  In addition, the median molecular fraction, $f$, assuming $b$ = 2 km s$^{-1}$, is $\sim$3.7 times smaller than that of the N08 sample.  Removing the caveat that the Magellan sample contains a mix of resolutions and considering only the MagE data, the median upper limit, N(\htwo ) = 10$^{17.7}$ cm$^{-2}$ (assuming $b$= 2 km s$^{-1}$), is $\approx$0.5\,dex below the median N(\htwo ) of the detections in the N08 sample, N(\htwo ) = 10$^{18.16}$ cm$^{-2}$.

We estimated the amount of \htwo\ we expected to find in the Magellan sample, under the assumption that the \htwo\ column density distribution found in N08 is an unbiased representation.  We find that the expected number of \htwo\ detections with N(\htwo ) $\ge$ 10$^{17.5}$ cm$^{-2}$ is 8.56 and the probability of detecting $\leq$ 1 system is 0.18\% from Poisson statistics.  As a result we conclude that the Magellan survey detects too few \htwo\ absorbers with high column densities, N(\htwo ) $\ge$ 10$^{17.5}$ cm$^{-2}$, if the N08 sample provides an unbiased representation of the N(\htwo ) distribution.  

While we confirm the reliability of the derived upper limits in Section~\ref{sec:testing}, we are also able to estimate the effects of a potential systematic underestimate of the upper limits (see analysis in Section~\ref{sec:expectedh2}). We find that by increasing all of the Magellan sample upper limits by 0.6\,dex, the number of expected detections with N(\htwo) $\ge$ 10$^{17.5}$ cm$^{-2}$ is \expectedtwosigma , with Posisson probability of $\le$1 of \probltonetwosigma \%, still in tension with the results of N08.  Indeed, in order to bring our detection of a single N(\htwo) $\ge$ 10$^{17.5}$ cm$^{-2}$ system into 95\% confidence agreement with N08, we would need to increase the Magellan upper limits by 1.8\,dex, a very large factor which is clearly inconsistent with our tests in detecting previously known H$_2$ systems in spectra of similar resolution and SNR as the Magellan sample.

We show that the \dlas\ with weak upper limits near N(\htwo ) $\sim$ 10$^{18}$ cm$^{-2}$ are generally not likely to be \htwo -bearing but rather have weak limits because of low SNR or limited spectral coverage.  However, if we include the 3 \dlas\ that may contain significant \htwo , the detection rate increases to $\sim$5\, \%.  This detection rate is still smaller, albeit at a lower confidence level ($\sim$89\, \% confidence), than the rate of $10.3^{+3.1}_{-4.6}$\,\% found by N08.  Clearly, these 3 systems need to be followed up with higher resolution spectroscopy in order to measure or rule out the presence of \htwo . 

The results of our survey -- an unexpected paucity of strong \htwo\ absorption in a blind, uniformly selected \dla\ sample -- give rise to the question: If \dlas\ are indeed the reservoirs of neutral gas for star formation across cosmic time and multiple lines of evidence show that at least some level of star formation is taking place in \dlas , where is all of the \htwo ?  As suggested by \cite{zwaan06}, it seems likely that the majority of the \htwo\ in \dlas\ is in fact concentrated in highly over-dense regions with very low covering factors ($\sim$1 $-$ 6\,\%) such that the detection of strong \htwo\ absorption in any given \dla\ is relatively rare.

\section*{Acknowledgements}

R. A. J. gratefully acknowledges support from the STFC-funded Galaxy Formation and Evolution programme at the Institute of Astronomy, University of Cambridge and the NSF Astronomy and Astrophysics Postdoctoral Fellowship under award AST-1102683.  M. T. M. thanks the Australian Research Council for a QEII Research Fellowship (DP0877998) and Discovery Project grant (DP130100568).  Australian access to the Magellan Telescopes was supported through the Major National Research Facilities program and National Collaborative Research Infrastructure Strategy of the Australian Federal Government.

\small
\itemindent -0.48cm
\bibliography{regina}

\end{document}

%% file: xavier_defs.tex
\def\kms{km~s$^{-1}$ }

\def\s-1{s$^{-1}$}
\def\Hz-1{Hz$^{-1}$}

\def\sci#1{{\; \times \; 10^{#1}}}

\def\l1l2{\lambda_2 \; {\rm and} \; \lambda_1}

\def\eht1{{\rm \hat e_1}}

\def\d3x{d^3 x}

\def\perd{\;\;\; .}
\def\delv{\Delta v}

\def\cm#1{\, {\rm cm^{#1}}}

%% file: MNRAS_tab_summ_lls.tex
\begin{table}
\centering
\caption{ DLAs with an Intervening LLS Rendering H$_2$ Unobservable.}
{\scriptsize \begin{tabular}{ccc}
\hline\hline
Quasar & $z_{abs}$ & $z_{LLS}$\\
\hline\hline
J0011$+$1446&3.4522&3.994\\
J0338$-$0005&2.2297&2.747\\
J0954$+$0915&2.4420&3.153\\
J1019$+$0825&2.3158&2.965\\
J1037$+$0910&2.8426&3.630\\
J1106$+$0816&3.2240&4.061\\
J1133$+$1305&2.5975&3.298\\
J1155$+$0530&2.6077&3.327$^a$\\
J1220$+$0921&3.3090&4.110$^a$\\
J1309$+$0254&2.2450&2.925\\
J1330$+$0340&2.3215&2.682\\
J2122$-$0014&3.2064&4.001\\
J2154$+$1102&2.4831&3.141\\
\hline\hline 
\end{tabular}
}
\label{tab:lls}
\begin{spacing}{0.7}
{\scriptsize {\bf $^a$} Higher $z_{abs}$ \dla . } \\
\end{spacing}
\end{table}

%% file: h2tab_summ_MNRAS1.tex
\begin{table*}
\centering
\caption{DLA Sample.}
{\scriptsize \begin{tabular}{lccccrrcccc}
\hline\hline
Quasar & $z_{abs}$ & N(H$_{2}^{J=1}$, b=1),frac$^b$ & N(H$_{2}^{J=1}$, b=2),frac$^b$ & N(H$_{2}^{J=1}$, b=10),frac$^b$ & wave$_{min}^{c}$ & $\Delta$$\lambda$$^d$ & med(S/N)$^e$ & med(S/N)$_{H2}$$^f$ & pixels$^g$ & I$^h$ \\  
\hline\hline
J0011+1446$^a$&3.6175&$<17.90, -2.29$&$<17.90, -2.29$&$<15.50, -4.69$&1073.08& 39.92&  32&  14&          13& 1\\
J0013+1358&3.2812&$<17.70, -3.29$&$<17.90, -3.09$&$<15.30, -5.69$& 948.59&164.41&  22&  18&          95& 1\\
J0035-0918&2.3401&$<17.70, -2.29$&$<17.50, -2.49$&$<15.30, -4.69$& 972.50&140.50&  18&  14&         114& 4\\
J0124+0044&3.0777&$<14.90, -4.89$&$<14.40, -5.39$&$<14.50, -5.29$&1023.57& 89.43&  28&  20&          16& 3\\
J0127+1405&2.4416&$<18.30, -1.55$&$<18.30, -1.55$&$<15.90, -3.94$&1012.63&100.37&  21&  12&          74& 1\\
J0139-0824&2.6773&$<14.50, -5.69$&$<14.10, -6.09$&$<12.05, -8.14$&1039.15& 73.85&  12&  12&          47& 3\\
J0211+1241&2.5947&$<18.30, -1.79$&$<18.30, -1.79$&$<15.90, -4.19$&1060.94& 52.06&  18&  11&          38& 1\\
J0234-0751&2.3181&$<18.10, -2.24$&$<18.10, -2.24$&$<15.70, -4.64$&1020.01& 92.99&  25&  12&         102& 1\\
J0239-0038&3.0185&$<17.10, -2.74$&$<17.10, -2.74$&$<14.70, -5.14$& 935.53&177.47&  21&  18&         172& 1\\
J0255+0048&3.2540&$<18.10, -2.09$&$<18.10, -2.09$&$<15.50, -4.69$&1057.83& 55.17&  23&  19&          10& 1\\
J0255+0048&3.9146&$<17.70, -3.09$&$<17.70, -3.09$&$<15.10, -5.69$& 915.64&197.36&  23&  24&          82& 1\\
J0912+0547&3.1236&$<16.90, -2.89$&$<16.50, -3.29$&$<14.50, -5.29$& 922.77&190.23&  21&  17&         137& 1\\
J0927+0746&2.3104&$<16.50, -3.79$&$<15.10, -5.19$&$<14.50, -5.79$&1022.65& 90.35&  38&  15&          35& 1\\
J0942+0422&2.3067&$<13.80, -5.99$&$<13.80, -5.99$&$<14.10, -5.69$&1020.52& 92.48&  36&  25&          99& 4\\
J0949+1115&2.7584&$<18.30, -2.14$&$<18.10, -2.34$&$<15.90, -4.54$&1090.10& 22.90&  24&  11&          20& 1\\
J1004+0018&2.5400&$<18.50, -2.09$&$<18.50, -2.09$&$<16.70, -3.89$&1012.19&100.81&  23&  14&          80& 1\\
J1004+0018&2.6855&$<17.70, -3.04$&$<17.50, -3.24$&$<15.10, -5.64$& 972.23&140.77&  23&  15&         106& 1\\
J1004+1202&2.7997&$<17.50, -2.94$&$<17.30, -3.14$&$<14.90, -5.54$& 949.51&163.49&  31&  16&         158& 1\\
J1020+0922&2.5931&$<17.90, -3.04$&$<17.90, -3.04$&$<15.50, -5.44$&1063.25& 49.75&  31&  11&          22& 1\\
J1022+0443&2.7416&$<16.90, -3.19$&$<16.30, -3.79$&$<14.70, -5.39$&1004.30&108.70&  22&  16&          40& 1\\
J1023+0709&3.3777&$<17.10, -2.79$&$<16.70, -3.19$&$<14.70, -5.19$& 937.01&175.99&  22&  18&         109& 1\\
J1029+1356&2.9938&$<17.30, -2.89$&$<17.30, -2.89$&$<14.70, -5.49$& 949.14&163.86&  20&  17&          92& 1\\
J1032+0149&2.2100&$<17.90, -2.04$&$<17.90, -2.04$&$<15.50, -4.44$&1018.00& 95.00&  32&  16&         103& 1\\
J1040-0015&3.5450&$<17.90, -2.34$&$<17.90, -2.34$&$<15.50, -4.74$&1049.50& 63.50&  24&  18&          20& 1\\
J1042+0117&2.2667&$<15.50, -4.74$&$<14.70, -5.54$&$<14.50, -5.74$&1005.43&107.57&  14&  11&         159& 3\\
J1048+1331&2.9196&$<17.50, -2.64$&$<17.50, -2.64$&$<14.90, -5.24$& 925.57&187.43&  23&  19&         154& 1\\
J1057+0629&2.4995&$<15.30, -4.69$&$<14.50, -5.49$&$<14.70, -5.29$&1068.87& 44.13&  13&  12&           2& 3\\
J1100+1122&3.7559&$<18.50, -1.74$&$<18.50, -1.74$&$<17.10, -3.14$&1051.34& 61.66&  45&  42&           6& 2\\
J1100+1122&4.3949&$<16.90, -3.99$&$<16.50, -4.39$&$<14.50, -6.39$& 885.52&227.48&  45&  49&           8& 2\\
J1108+1209&3.5454&$<13.70, -6.54$&$<13.60, -6.64$&$<13.90, -6.34$& 946.89&166.11&  26&  35&          83& 3\\
J1108+1209&3.3963&$<14.10, -6.04$&$<14.00, -6.14$&$<14.30, -5.84$& 979.00&134.00&  26&  23&          39& 3\\
J1111+0714&2.6820&$<17.50, -2.59$&$<17.30, -2.79$&$<14.70, -5.39$& 925.47&187.53&  30&  18&         137& 1\\
J1111+1332&2.3822&$<17.30, -2.64$&$<17.10, -2.84$&$<14.70, -5.24$& 954.32&158.68&  41&  19&         125& 1\\
J1111+1332&2.2710&$<17.70, -2.29$&$<17.50, -2.49$&$<15.10, -4.89$& 986.76&126.24&  41&  17&          87& 1\\
J1111+1336&3.2004&$<14.30, -6.34$&$<14.00, -6.64$&$<14.00, -6.64$& 910.15&202.85&  15&  12&         220& 3\\
J1111+1442&2.5996&$<15.50, -5.34$&$<14.70, -6.14$&$<14.70, -6.14$&1028.94& 84.06&  12&  12&          99& 3\\
J1133+0224&3.9155&$<17.30, -2.74$&$<17.10, -2.94$&$<14.70, -5.34$& 906.10&206.90&  20&  18&          74& 1\\
J1140+0546&2.8847&$<17.70, -2.14$&$<17.70, -2.14$&$<15.10, -4.74$& 985.81&127.19&  17&  14&         110& 1\\
J1142-0012&2.2578&$<17.90, -1.99$&$<17.90, -1.99$&$<15.30, -4.59$&1002.86&110.14&  27&  15&          82& 1\\
J1151+0552&2.9287&$<16.70, -3.64$&$<15.90, -4.44$&$<14.50, -5.84$& 917.41&195.59&  30&  26&          83& 1\\
J1153+1011&3.7950&$<17.10, -3.74$&$<15.70, -5.14$&$<14.50, -6.34$& 949.48&163.52&  24&  25&          42& 1\\
J1153+1011&3.4695&$<17.90, -2.34$&$<17.90, -2.34$&$<15.30, -4.94$&1030.76& 82.24&  24&  21&          27& 1\\
J1155+0530&3.3261&$<13.70, -6.84$&$<13.70, -6.84$&$<14.00, -6.54$& 913.55&199.45&  29&  28&         126& 3\\
J1201+0116&2.6852&$<13.00, -7.49$&$<13.00, -7.49$&$<13.60, -6.89$& 949.78&163.22&  26&  25&          91& 4\\
J1208+0043&2.6084&$<17.50, -2.44$&$<17.50, -2.44$&$<14.90, -5.04$& 945.80&167.20&  26&  16&         200& 1\\
J1211+0422&2.3766&$<13.70, -6.44$&$<13.60, -6.54$&$<14.00, -6.14$&1019.62& 93.38&  17&  16&         273& 4\\
J1211+0902&2.5835&$<14.50, -6.29$&$<14.30, -6.49$&$<14.70, -6.09$&1055.56& 57.44&  20&  16&          12& 3\\
J1223+1034&2.7194&$<16.70, -3.24$&$<15.70, -4.24$&$<12.05, -7.89$& 912.75&200.25&  22&  17&         108& 1\\
J1226+0325&2.5078&$<17.50, -2.94$&$<17.30, -3.14$&$<14.90, -5.54$& 883.74&229.26&  19&  13&          18& 1\\
J1226-0054&2.2903&$<18.50, -1.70$&$<18.50, -1.70$&$<16.50, -3.69$&1054.96& 58.04&  27&  13&          45& 1\\
J1228-0104$^a$&2.2625&$<14.50, -5.39$&$<14.40, -5.49$&$<14.70, -5.19$&1013.43& 99.57&  17&   7&         175& 3\\
J1233+1100&2.7924&$<17.90, -2.19$&$<17.90, -2.19$&$<15.50, -4.59$& 971.86&141.14&  16&  14&         225& 1\\
J1233+1100&2.8206&$<16.70, -3.14$&$<15.90, -3.94$&$<14.50, -5.34$& 964.69&148.31&  16&  14&         161& 1\\
J1240+1455&3.0242&$<14.90, -4.89$&$<14.30, -5.49$&$<14.30, -5.49$& 949.48&163.52&  14&  11&         236& 3\\
J1246+1113&3.0971&$<17.30, -2.64$&$<17.10, -2.84$&$<14.70, -5.24$& 930.35&182.65&  19&  17&         138& 1\\
J1253+1147&2.9443&$<14.50, -5.39$&$<14.20, -5.69$&$<14.30, -5.59$& 969.40&143.60&  15&  13&         203& 3\\
J1253+1306&2.9812&$<17.70, -2.39$&$<17.70, -2.39$&$<15.10, -4.99$&1022.59& 90.41&  23&  15&          17& 1\\
J1257-0111&4.0209&$<14.90, -4.94$&$<14.10, -5.74$&$<13.90, -5.94$& 899.47&213.53&  30&  33&          39& 1\\
J1304+1202&2.9133&$<17.30, -2.74$&$<17.10, -2.94$&$<14.70, -5.34$& 929.91&183.09&  19&  14&         175& 1\\
J1304+1202&2.9288&$<17.50, -2.34$&$<17.30, -2.54$&$<14.90, -4.94$& 926.24&186.76&  19&  14&         238& 1\\
J1306-0135$^a$&2.7730&$<17.10, -2.99$&$<17.10, -2.99$&$<15.70, -4.39$&1022.94& 90.06&   6&   6&          99& 1\\
J1317+0100&2.5365&$<17.10, -3.94$&$<16.50, -4.54$&$<14.90, -6.14$& 963.41&149.59&  33&  16&         289& 2\\
J1337-0246&2.6871&$<17.50, -2.59$&$<17.30, -2.79$&$<14.90, -5.19$& 998.73&114.27&  22&  18&          84& 2\\
J1339+0548&2.5851&$<14.20, -5.89$&$<14.00, -6.09$&$<14.20, -5.89$& 970.62&142.38&  20&  17&         101& 3\\
J1340+1106&2.7958&$<13.25, -7.09$&$<13.25, -7.09$&$<12.65, -7.69$& 910.43&202.57&  29&  24&         301& 3\\
J1344-0323&3.1900&$<17.90, -2.54$&$<17.70, -2.74$&$<15.30, -5.14$& 971.37&141.63&  12&  11&          90& 1\\
J1353-0310&2.5600&$<16.10, -3.94$&$<15.10, -4.94$&$<14.90, -5.14$&1049.22& 63.78&  12&  12&          71& 3\\
J1358+0349&2.8516&$<16.50, -3.39$&$<15.50, -4.39$&$<14.40, -5.49$& 915.28&197.72&  27&  25&         439& 2\\
J1402+0117&2.4295&$<18.30, -1.50$&$<18.30, -1.50$&$<15.70, -4.09$&1068.44& 44.56&  22&  15&          44& 2\\
J1450-0117&3.1901&$<17.70, -2.99$&$<17.70, -2.99$&$<15.10, -5.59$& 985.84&127.16&  16&  12&          74& 1\\
J1453+0023&2.4440&$<17.70, -2.14$&$<17.70, -2.14$&$<15.30, -4.54$& 997.65&115.35&  20&  12&         156& 1\\
J1550+0537&2.4159&$<18.10, -2.04$&$<18.10, -2.04$&$<15.70, -4.44$&1075.46& 37.54&  25&  10&          26& 1\\
\hline\hline
\end{tabular}
}
\label{tab:h2sample}
\begin{spacing}{0.7}
{\scriptsize {\bf$^{a}$} System in which SNR threshold in H$_2$ region was set to 5 rather than 10, i.e. there is a LLS but it did not obliterate ALL H$_2$ lines.} \\
{\scriptsize {\bf$^{b}$} Log of the molecular fraction calculated as defined in text.} \\
{\scriptsize {\bf$^{c}$} Lowest rest-frame wavelength searched for \htwo . This cutoff was determined by the requirement that S/N $\geq$ 10.}
{\scriptsize {\bf$^{d}$} \htwo\ upper limit was determined over this range in Angstroms.} \\
{\scriptsize {\bf$^{e}$} Median S/N per pixel over the spectral region used (S/N $>$ 10)} \\
{\scriptsize {\bf$^{f}$} Median S/N per pixel over the \htwo\ spectral region used, sometimes S/N threshold was set to 5 to have coverage.} \\
{\scriptsize {\bf$^{g}$} Number of data pixels used to constrain upper limit assuming b = 2 km s$^{-1}$.} \\
{\scriptsize {\bf$^{h}$} Instrument Used: 1 = MagE (FWHM$\sim$71 km s$^{-1}$), 2 = XShooter (FWHM$\sim$59 km s$^{-1}$) , 3 = UVES (FWHM$\sim$8 km s$^{-1}$), 4 = HIRES (FWHM$\sim$8 km s$^{-1}$)} \\
\end{spacing}
\end{table*}

%% file: h2tab_summ_MNRAS2.tex
\begin{table*}
\contcaption{DLA Sample.}
\centering
{\scriptsize \begin{tabular}{lccccrrcccc}
\hline\hline
Quasar & $z_{abs}$ & N(H$_{2}^{J=1}$, b=1),frac$^b$ & N(H$_{2}^{J=1}$, b=2),frac$^b$ & N(H$_{2}^{J=1}$, b=10),frac$^b$ & wave$_{min}^{c}$ & $\Delta$$\lambda$$^d$ & med(S/N)$^e$ & med(S/N)$_{H2}$$^f$ & pixels$^g$ & I$^h$ \\  
\hline\hline
J2036-0553&2.2804&$<14.30, -6.39$&$<14.10, -6.59$&$<14.50, -6.19$&1046.40& 66.60&  12&  11&         231& 4\\
J2049-0554&2.6828&$<16.70, -3.09$&$<16.50, -3.29$&$<14.50, -5.29$& 975.67&137.33&  33&  19&          45& 1\\
J2141+1119&2.4263&$<17.70, -2.09$&$<17.70, -2.09$&$<15.10, -4.69$& 999.66&113.34&  24&  13&         171& 1\\
J2222-0946&2.3544&$<14.30, -5.84$&$<14.10, -6.04$&$<14.40, -5.74$&1054.10& 58.90&  16&  11&         148& 4\\
J2238+0016&3.3654&$<16.90, -3.14$&$<16.50, -3.54$&$<14.70, -5.34$& 920.42&192.58&  35&  26&         131& 1\\
J2238-0921&2.8691&$<16.30, -3.84$&$<15.10, -5.04$&$<14.50, -5.64$& 966.68&146.32&  32&  23&          59& 1\\
J2241+1225&2.4175&$<17.30, -3.29$&$<17.10, -3.49$&$<14.90, -5.69$& 987.50&125.50&  31&  16&          86& 1\\
J2241+1352&4.2833&$<16.10, -4.44$&$<15.10, -5.44$&$<14.20, -6.34$& 919.33&193.67&  28&  32&          33& 1\\
J2315+1456&3.2730&$<17.30, -2.49$&$<16.70, -3.09$&$<14.90, -4.89$& 898.56&214.44&  29&  29&          93& 1\\
J2334-0908&3.0572&$<13.30, -6.59$&$<13.25, -6.64$&$<13.70, -6.19$& 941.44&171.56&  51&  54&          43& 3\\
J2343+1410&2.6768&$<15.30, -4.69$&$<14.90, -5.09$&$<15.10, -4.89$&1080.57& 32.43&  11&  11&         111& 4\\
J2348-1041&2.9979&$<13.80, -6.24$&$<13.70, -6.34$&$<13.90, -6.14$& 902.91&210.09&  17&  18&         161& 4\\
J2350-0052&2.6147&$<13.60, -7.14$&$<13.50, -7.24$&$<13.90, -6.84$& 988.24&124.76&  32&  35&          40& 3\\
\hline\hline
\end{tabular}
}
\begin{spacing}{0.7}
{\scriptsize {\bf$^{a}$} System in which SNR threshold in H$_2$ region was set to 5 rather than 10, i.e. there is a LLS but it did not obliterate ALL H$_2$ lines.} \\
{\scriptsize {\bf$^{b}$} Log of the molecular fraction calculated as defined in text.} \\
{\scriptsize {\bf$^{c}$} Lowest rest-frame wavelength searched for \htwo . This cutoff was determined by the requirement that S/N $\geq$ 10.}
{\scriptsize {\bf$^{d}$} \htwo\ upper limit was determined over this range in Angstroms.} \\
{\scriptsize {\bf$^{e}$} Median S/N per pixel over the spectral region used (S/N $>$ 10)} \\
{\scriptsize {\bf$^{f}$} Median S/N per pixel over the \htwo\ spectral region used, sometimes S/N threshold was set to 5 to have coverage.} \\
{\scriptsize {\bf$^{g}$} Number of data pixels used to constrain upper limit assuming b = 2 km s$^{-1}$.} \\
{\scriptsize {\bf$^{h}$} Instrument Used: 1 = MagE (FWHM$\sim$71 km s$^{-1}$), 2 = XShooter (FWHM$\sim$59 km s$^{-1}$) , 3 = UVES (FWHM$\sim$8 km s$^{-1}$), 4 = HIRES (FWHM$\sim$8 km s$^{-1}$)} \\
\end{spacing}
\end{table*}

%% file: MNRAS_likelihood.tex
\begin{table}
\centering
\caption{ Constrained transition fraction for DLAs with weak N(\htwo ) upper limits.}
{\scriptsize \begin{tabular}{ccc}
\hline\hline
Quasar & $z_{abs}$ & C$^a$ \\
\hline\hline
&Smoothed UVES\\
Q0013$-$0029&1.973&0.09\\
Q0027$-$1836&2.402&0.31\\
Q0347$-$3819&3.025&0.36\\
Q0405$-$4418&2.595&0.33\\
Q0528$-$2505&2.811&0.35\\
Q0551$-$3638&1.962&0.02\\
Q0642$-$5038&2.659&0.50\\
Q1232$+$0815&2.338&0.16\\
Q1441$+$2737&4.224&0.13\\
Q1444$+$0126&2.087&0.10\\
Q2318$-$1107&1.989&0.10\\
Q2343$+$1232&2.431&0.39\\
\hline
&Magellan Sample\\
J0127$+$1405&2.4416&0.13\\
J0211$+$1241&2.5947&0.04\\
J0234$-$0751&2.3181&0.15\\
J0255$+$0048&3.2540&0.04\\
J0949$+$1115&2.7584&0.03\\
J1004$+$0018&2.5400&0.13\\
J1100$+$1122&3.7559&0.01\\
J1226$-$0054&2.2903&0.07\\
J1402$+$0117&2.4295&0.04\\
J1550$+$0537&2.4159&0.04\\
\hline\hline 
\end{tabular}
}
\label{tab:likelihood}
\begin{spacing}{0.7}
{\scriptsize {\bf $^a$} Constrained transition fraction, $C = \frac{constrained}{total}$, as explained in the text.} \\
\end{spacing}
\end{table}